\documentclass[aps,prd,preprint,showpacs,groupaddress,nofootinbib]{revtex4}
\usepackage{color}
\usepackage{graphicx}
\usepackage{tabularx}
\usepackage{amsmath}
\usepackage{amssymb}
\usepackage{datetime}

\usepackage[left]{lineno}

\begin{document}

\title{Exploration of decaying dark matter   in \\ a  left-right symmetric model}

\author{Wan-Lei Guo}
\email[Email: ]{guowl@itp.ac.cn}

\author{Yue-Liang Wu}
\email[Email: ]{ylwu@itp.ac.cn}

\author{Yu-Feng Zhou}
\email[Email: ]{yfzhou@itp.ac.cn}

\affiliation{ Kavli Institute for Theoretical Physics China, \\
Key Laboratory of Frontiers in Theoretical Physics, \\
Institute of Theoretical Physics, Chinese Academy of Science, Beijing 100190,
China}


\begin{abstract}
  $SU(2)_L$ triplet scalars appear in models motivated for the
  left-right symmetry, neutrino masses and dark matter (DM), etc..  If
  the triplets are the main decay products of the DM particle, and
  carry nonzero lepton numbers, they may decay dominantly into lepton
  pairs, which can naturally explain the current experimental results
  reported by PAMELA and Fermi-LAT or ATIC.  We discuss this
  possibility in an extended left-right symmetric model in which the
  decay of DM particle is induced by tiny soft charge-conjugation
  ($C$) violating interactions, and calculate the spectra for
  cosmic-ray positrons, neutrinos and gamma-rays.  We show that the DM
  signals in the flux of high energy neutrinos can be significantly
  enhanced, as the triplets couple to both charged leptons and
  neutrinos with the same strength. In this scenario, the predicted
  neutrino-induced muon flux can be several times larger than the case
  in which DM particle only directly decays into charged leptons.  In
  addition, the charged components of the triplet may give an extra
  contribution to the high energy gamma-rays through internal
  bremsstrahlung process, which depends on the mass hierarchy between
  the DM particle and the triplet scalars.
\end{abstract}


\maketitle

\section{Introduction}\label{Sec:Introduction}

Recently, the PAMELA satellite experiment has observed an excess in
the positron fraction from 10 to 100 GeV \cite{PAMELA}, which
confirmed the previous hints from HEAT \cite{HEAT}, CAPRICE
\cite{CAPRICE} and AMS-01 \cite{AMS}.  The reports from ATIC balloon
experiment showed a rapid rise of the total electron and positron
flux at a range 300-800 GeV\cite{ATIC} with a peak at around 600 GeV,
which agreed with the PPB-BETS results\cite{PPB}.  More recently, the
Fermi-LAT \cite{FermiElectron} and the HESS \cite{HESS} Cherenkov
experiments have released their results for the sum of electron and
positron flux. Although they did not fully confirm the previous
results reported by the ATIC experiment, both the measurements
indicated excesses over the expected background, which suggests the
presence of extra sources for the $e^\pm$ spectra.  Besides the
astrophysical explanations for $e^\pm$ anomalies by some nearly
sources like pulsars and supernova remnants, the dark matter (DM)
annihilation or decay is one of the most interesting explanation from
particle physics.

The PAMELA, ATIC, and Fermi-LAT anomalies 
may be a consequence of DM particle annihilating mainly into lepton
final states. However, if the dark matter is generated thermally, the
annihilation cross section obtained from the observed relic density is
significantly lower than that required by the current data. One has to
resort to a large boost factor (about $100-1000$) to explain the
observed large positron flux. Note that the most probable boost factor
from the clumpiness of DM structures is estimated to be less than
$\sim10-20$ \cite{Structure}. The large boost factor may come from
nonperturbative Sommerfeld enhancement \cite{Sommerfeld} or
Breit-Wigner enhancement \cite{BW,Bi:2009uj,Gogoladze:2009gi}.  An other possibility is that the
DM particle may slightly decay, and dominantly decay into leptons
\cite{Decay}. In this case the bound from the DM relic density is
relaxed. The current data require that the lifetime of DM particle
should be of the order $\mathcal{O}(10^{26}$s).

If the mass of DM particle is very heavy, it may first decay  into
some lighter intermediate  states rather than directly into
standard model (SM) particles. An interesting intermediate state
might be $SU(2)_L$ triplets, as it is known  in many models  that they
may carry nonzero lepton numbers, and do not couple to quarks directly.
This leptophilic feature can naturally account for the PAMELA
results on both the excesses of positrons and the absence of large
anti-proton flux.

The stability of the DM particle is usually protected by imposing
extra discrete symmetries. In our  previous work \cite{Guo:2008si},
we have shown that the fundamental symmetries of quantum field
theory such as parity ($P$) and charge-conjugation ($C$) can be used
to stabilize the DM particle. We have explicitly demonstrated  that
in a left-right (LR) symmetry model with $P$ and $CP$ only broken
spontaneously, a gauge singlet scalar with odd $CP$ parity can be
automatically stable without imposing any extra discrete symmetries.

Motivated by the recent experimental results, we consider in this work
the possibility of DM particle decay by adding soft $C-$violating
interactions into our previous  model, which will result in the
decay of DM particle with  $SU(2)_L$ triplet scalars as intermediate
states. The decays of the triplets into the SM gauge boson  pairs
$W^\pm W^\pm$, $W^\pm Z^0$, $Z^0 Z^0$ as well as the SM-like Higgs bosons
$h^0h^0$ are all suppressed by the smallness of the vacuum
expectation value (VEV) of the left-handed triplet as required by
the tiny neutrino masses. Thus the triplets  decaying into 
quarks indirectly through these states are strongly suppressed.

If the reported positron excess indeed originates from DM particle
decay through the intermediate $SU(2)_L$ triplets, the possible
associating signals such as the high energy neutrinos and gamma-rays
are expected. This is because the neutral and singly charged
components of the triplets couple to neutrinos in the same way as the
doubly charged components couple to charged leptons. The final state
leptons may generate high energy gamma-rays through inverse Compton
scatterings (ICS) and final state radiations (FSR), which is common to
many DM models.  In this model the doubly and singly charged
components of the triplet can also produce extra gamma-rays through
virtual internal bremsstrahlung (VIB), which gives additional
contributions to the gamma-ray spectrum at very high energy region.

In this work we first show how the DM particle decay caused by the
soft $C$-violating interactions can naturally explain the observed
excess of the positron and electron excesses, and then focus on the
cosmic-ray signals associated with DM particle decay through
intermediate $SU(2)_L$ triplets in a LR symmetric model with two Higgs
bidoublets \cite{Wu:2007kt,Guo:2008si}.  After exploring the typical
parameters which can explain the current PAMELA as well as Fermi-LAT
or ATIC data, we present predictions for the flux of neutrinos and
gamma-rays.  We find that the predicted neutrino-induced muon flux can
be significantly larger than the case in which DM particle only
directly decays into charged leptons.  The energy spectrum of diffuse
gamma-rays can be enhanced by the VIB processes from internal charged
triplets as well. Although the analysis is done in a particular model,
the conclusions are generally valid for other models which involve
$SU(2)_L$ triplets as intermediate states, such as the DM models in
connection to the type II seesaw mechanism for neutrino masses (see,
e.g. Ref. \cite{Gogoladze:2009gi}).

This paper is organized as follows: in Sec. \ref{Sec:Model}, we
discuss the main features of the model and focus on the decay of DM
particle induced by the soft $C$-breaking terms. In Sec.
\ref{Sec:Propagation}, we calculate the DM decay signals which
includes the positron fraction, total flux of electron and positron,
neutrino-induced muon flux and high energy gamma-ray flux. The
results are summarized  in Sec. \ref{Sec:Conclusion}.

\section{Decaying dark matter in a LR symmetric model }\label{Sec:Model}

We begin with a brief review of the LR model with two Higgs
bidoublets described in Ref. \cite{Wu:2007kt,Guo:2008si}.  The model
is a simple extension to the minimal LR model \cite{minLR}, which is
based on the gauge group $SU(2)_L\otimes SU(2)_R\otimes U(1)_{B-L}$.
The left- and right-handed fermions belong to the $SU(2)_L$ and
$SU(2)_R$ doublets  respectively.
The Higgs sector  contains two Higgs bidoublets $\phi$
(2,$2^{*}$,0), $\chi$ (2,$2^{*}$,0), a left(right)-handed Higgs
triplet $\Delta_{L(R)}$ (3(1),1(3),2), and a gauge singlet
$S$(0,0,0) with the following flavor contents
\begin{eqnarray}
\phi  = \left ( \begin{matrix} \phi_1^0 & \phi_1^+ \cr \phi_2^- &
\phi_2^0 \cr  \end{matrix} \right ) , \;
\chi  = \left ( \begin{matrix} \chi_1^0 & \chi_1^+ \cr \chi_2^- &
\chi_2^0 \cr  \end{matrix} \right ) , \;
\Delta_{L,R}  = \left (
\begin{matrix} \delta_{L,R}^+/\sqrt{2} & \delta_{L,R}^{++} \cr
\delta_{L,R}^{0} & -\delta_{L,R}^{+}/\sqrt{2} \cr \end{matrix}
\right ) , \;
S=\frac{1}{\sqrt2}(S_\sigma+i S_D) .
\nonumber\\
\end{eqnarray}
The two triplets $\Delta_{L,R}$ are responsible for breaking the
left-right symmetry at high energy scale and generating  small
neutrino masses through the seesaw mechanism.  The introduction of
Higgs bidoublets $\phi$ and $\chi$  accounts for the electroweak
symmetry breaking and overcome the fine-tuning problem in generating
the spontaneous $CP$ violation in the minimal LR model. Meanwhile it
also relaxes the severe low energy phenomenological constraints. The
gauge singlet $S$ is relevant to the DM candidate.

The kinematic terms for the scalar fields are given by
\begin{align}
\mathcal{L}=
\mbox{Tr}[(D_\mu\phi)^\dagger (D^\mu\phi)]
+\mbox{Tr}[(D_\mu \Delta_L)^\dagger(D^\mu \Delta_L)] +(\phi
\leftrightarrow \chi, L \leftrightarrow R) \;,
\end{align}
where
\begin{align}
D_\mu \phi&=\partial_\mu \phi -ig \frac{\tau}{2}W_{L\mu}\phi +ig
\phi \frac{\tau}{2}W_{R\mu}\;,
\nonumber\\
D_\mu \Delta_{L}&=\partial_\mu \Delta_{L}-ig
[\frac{\tau}{2}W_{L\mu}\ ,\ \Delta_{L}]-i g' B_\mu \Delta_{L} \;.
\end{align}
Due to constraints from low energy phenomenology, the mixing between
$W_L$ and $W_R$ is rather small.  Thus the SM gauge boson $W^\pm$ is
mostly $W_L^\pm$.

In this work, we shall focus on the scalar sector which is relevant to the  DM particle.
Under the $P$ and $CP$ transformations,  the scalar fields transform as follows
\begin{eqnarray}
\begin{tabular}{lll}\hline\hline
  &  $P$  &   $CP$\\ \hline
$\phi$ \;   &  $\phi^\dagger$ \; &$\phi^*$
\\
$\chi$ \;   &  $\chi^\dagger$ \; &$\chi^*$
\\
$\Delta_{L(R)}$ \;  & $\Delta_{R(L)}$ \; & $\Delta_{L(R)}^*$
\\
$S$  &  $S$ &   $S^*$
\\ \hline\hline
\end{tabular}
\label{pcp}
\end{eqnarray}
The whole scalar potential contains two parts
\begin{align}
\mathcal{V}=\mathcal{V}_0+\mathcal{V}_1.
\end{align}
It is required that the dominant part $\mathcal{V}_0$ is both $P$- and
$CP$-invariant while a small part $\mathcal{V}_1$ contains the soft
$C$-violating interactions.  The most general form for $\mathcal{V}_0$
is given by \cite{Guo:2008si}
\begin{eqnarray}
-\mathcal{V}_{0}&=&\frac{1}{\sqrt{2}}\tilde\mu_0^3(S+S^*)-\tilde\mu_S^2SS^*-\frac{1}{4}\tilde\mu_{\sigma}^2(S+S^*)^2
+\sqrt{2}\tilde\mu_{\sigma S}(S+S^*)SS^*\nonumber\nonumber\\
&&+\frac{1}{6\sqrt{2}}\tilde\mu_{3\sigma}(S+S^*)^3
+\tilde\lambda_S(SS^*)^2-\frac{1}{4}\tilde\lambda_{\sigma S}(S+S^*)^2SS^*-\frac{1}{16}\tilde\lambda_{\sigma}(S+S^*)^4\nonumber\\
&&+\sum_{i=1}^5\left[-\frac{1}{\sqrt{2}}\tilde\mu_{i,\sigma}(S+S^*)+\tilde\lambda_{i,S}SS^*
-\frac{1}{4}\tilde\lambda_{i,\sigma}(S+S^*)^2\right]O_i \, ,
\label{VS}
\end{eqnarray}
where
\begin{eqnarray}
O_1&=&{\rm{Tr}}(\Delta_L^{\dag}\Delta_L+\Delta_R^{\dag}\Delta_R),
O_2={\rm{Tr}}(\phi^{\dag}\phi),
O_3={\rm{Tr}}(\phi^{\dag}\tilde\phi+\tilde\phi^{\dag}\phi)
\nonumber\\
O_4&=&{\rm{Tr}}(\chi^{\dag}\chi),
O_5={\rm{Tr}}(\chi^{\dag}\tilde\chi+\tilde\chi^{\dag}\chi) \;.
\label{operators}
\end{eqnarray}
Due to the $C$ and $CP$ symmetries, $\mathcal{V}_0$ only involves
combinations of $(S+S^*)$ and $SS^*$. The interactions containing
odd powers of $(S-S^*)$ are forbidden  as they are $P$-even but
$C$-odd. Furthermore, $(S-S^*)$ cannot mix with the Higgs multiplets
in $O_i$ because the five independent gauge-invariant combinations
$O_i (i=1,\dots,5)$ in Eq. (\ref{operators}) are both $P$- and
$C$-even. Other possible Higgs multiplet combinations such as
${\rm{Tr}}(\phi^{\dag}\tilde\phi-\tilde\phi^{\dag}\phi)$ and
${\rm{Tr}}(\Delta_L^{\dag}\Delta_L-\Delta_R^{\dag}\Delta_R)$ are
$P$-odd, thus cannot couple to $(S-S^*)$.  The terms proportional to
even powers of $(S-S^*)$ can be rewritten in terms of $(S+S^*)^2$
and $SS^*$. The absence of odd power term of $(S-S^*)$ means that 
$S_D=(S-S^*)/(i\sqrt{2})$ is a potential DM candidate.

The spontaneous symmetry breaking (SSB) scheme is to realize the
breaking pattern $SU(2)_L\otimes SU(2)_R\otimes U(1)_{B-L}\to
SU(2)_L\otimes U(1)_Y\to U(1)_{\mathrm{em}}$.  After the SSB, the Higgs
multiplets obtain nonzero VEVs
\begin{equation}
\langle\delta_{L,R}^0\rangle=\frac{v_{L,R}}{\sqrt2}
,\;
\langle\phi_{1,2}^0\rangle=\frac{\kappa_{1,2}}{\sqrt2}
\;
\mbox{ and }
\langle\chi_{1,2}^0\rangle=\frac{\xi_{1,2}}{\sqrt2} ,
\end{equation}
where $\kappa_{1,2}$, $\xi_{1,2}$, $v_L$ and $v_R$ are in general
complex, and $ \kappa \equiv \sqrt{|\kappa_1|^2 +
  |\kappa_2|^2+|\xi_1|^2 + |\xi_2|^2} \approx 246$ GeV represents the
electroweak symmetry breaking scale. The value of $v_R$ sets the scale
of LR symmetry breaking which is directly linked to the right-handed
gauge boson masses $M_{W_R}=g v_R/\sqrt{2}$.  It is required that
$S_D$ does not develop a VEV directly from $\mathcal{V}_0$, namely $CP$ is not
broken by the singlet sector. It follows that after the SSB, although
$P$ and $CP$ are both broken, there is a residual $Z_2$ type of
discrete symmetry on $S_D$ in the gauge singlet sector, which is
induced from the original $CP$ symmetry.

Constraints on the parameter space from the DM relic density  has
been discussed in Ref. \cite{Guo:2008si}. The main DM annihilation
channels are $S_DS_D\to W^\pm W^\mp, t\bar{t}$ and SM-like Higgs
boson $h^0 h^0$. Making use of the constrained parameter space, we
predicted the weakly interacting massive particle-nucleus elastic
scattering cross section.  The results show that the typical
scattering cross section is about one order of magnitude below the
current experimental upper bound. But in the enhanced Yukawa
coupling case, the resulting spin-independent cross section can
reach $10^{-44} \mathrm{cm}^2$ which is close to the improved bound set 
by the recent CDMS-II experiment \cite{Ahmed:2009zw}.

The tiny soft $C$-breaking terms in $\mathcal{V}_1$  may lead to the DM particle
decay. The most general $P$-even but $C$-odd terms with dimension
less than four are given by
\begin{align}
-\mathcal{V}_1=\mu_\epsilon(S-S^*)\left[ \sum_{i=1}^5 \zeta_i  O_i +\zeta_6 (S+S^*)^2+\zeta_7 (S-S^*)^2\right] \ .
\end{align}
In this case, the $S_D$ may decay dominantly into two lighter scalars
such as $\delta_L \delta_L$.  The lifetime of $S_D$ is estimated by
$1/\tau_D \approx \mu_\epsilon^2\sqrt{1-4m^2/m_{D}^2}/(16\pi m_{D})$ for
$m\ll m_{D}$, where $m_D$ and $m$ stand for the mass of $S_D$ and
final state scalars respectively.  In order to have a lifetime around
$\mathcal{O}(10^{26}\mathrm{s})$, the required value of $\mu_\epsilon$ is
around $10^{-23}$ GeV.

\begin{figure}[htb]
\includegraphics[width=0.25\textwidth]{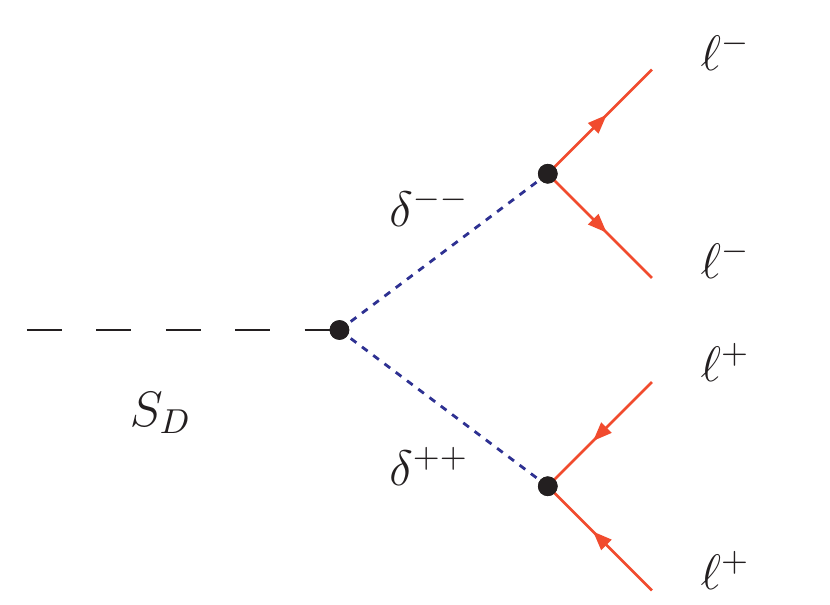}
\includegraphics[width=0.25\textwidth]{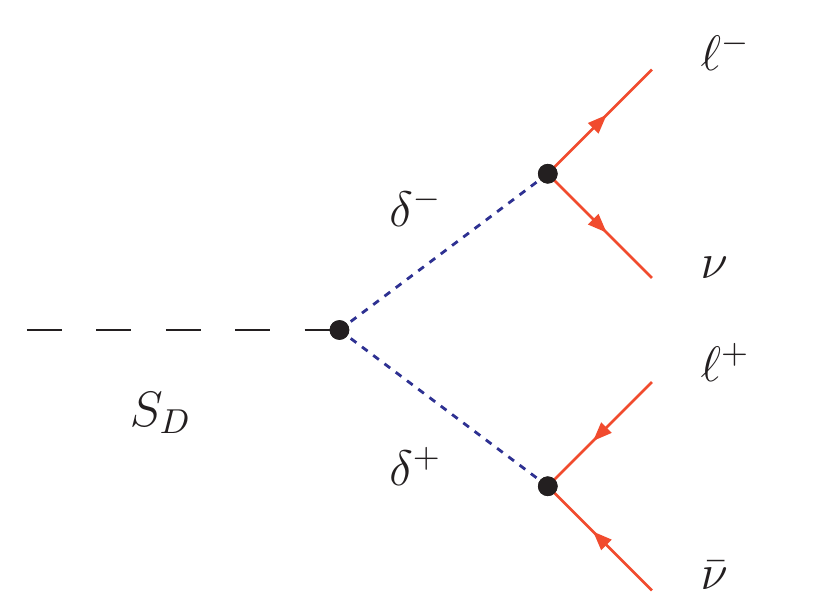}
\includegraphics[width=0.25\textwidth]{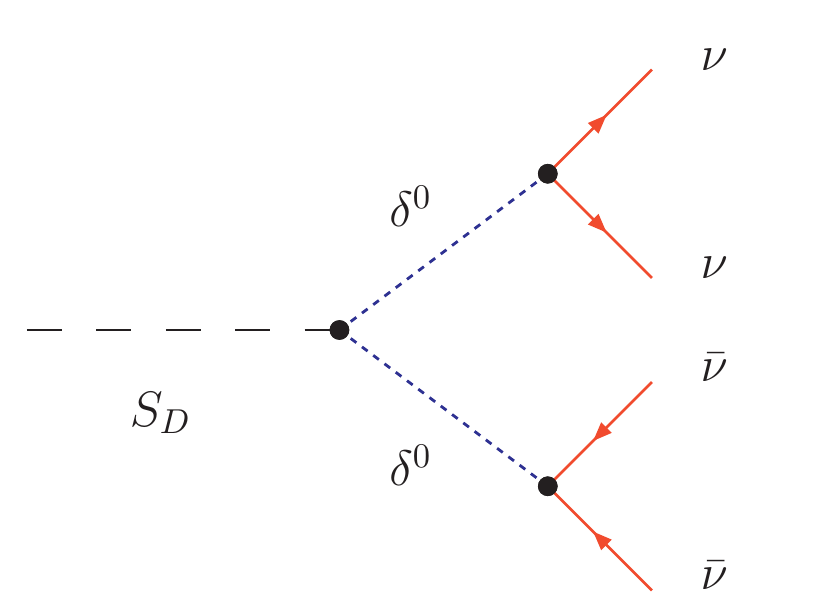}
\caption{Feynman diagrams for DM particle $S_D$ decaying into
charged leptons and neutrinos through intermediate $SU(2)_L$
triplets $\delta_L^{++}\delta_L^{--}$, $\delta_L^{+}\delta_L^{-}$ and $\delta_L^{0}\delta_L^{0}$.}
\label{fig:decay} 
\end{figure}

In this work we focus on the case in which
$\zeta_1 \gg \zeta_i \ (i=2,\dots,7)$ and the right-handed triplet
$\Delta _R$ is much heavier than that of $S_D$, such that $S_D$ will
decay dominantly into left-handed triplet $\Delta_L$ as shown in
Fig. \ref{fig:decay}.

The triplets with non-zero lepton number do not couple to quarks
directly.  The Yukawa interaction between the scalars and leptons is
\begin{align}
\mathcal{L}_{Y}=& -\bar{\ell}_{Li}
(h_{ij}\phi+\tilde{h}_{ij}\tilde\phi
+g_{ij}\chi+\tilde{g}_{ij}\tilde\chi)\ell_{Rj}
\nonumber\\
&-y_{ij} \left( \overline{\ell^{\mbox{ }c}_{Li}} i\tau_2 \Delta_L
\ell_{Lj} +\overline{\ell^{\mbox{ }c}_{Ri} } i\tau_2 \Delta_R
\ell_{Rj} \right) +h.c ,
\end{align}
which leads to the following seesaw formula for left-handed neutrino mass matrix
\begin{align}
  M_\nu=M_L-M_D \frac{1}{M_R} M_D^T \ 
\end{align}
with $(M_{L(R)})_{ij}=y_{ij} v_{L(R)}/\sqrt{2}$ and
$(M_D)_{ij}=(h_{ij}\kappa_1+\tilde{h}_{ij}\kappa_2^*
+g_{ij}\xi_1+\tilde{g}_{ij}\xi_2^*)/\sqrt{2}$.  The upper bound
for $v_L$ from neutrino masses is around a few eV, and that from the $\rho$-parameter in electroweak
precision measurements is about 1 GeV.  The absence of $W_R$ from collider searches sets a
lower bound of $v_R \gtrsim \mathcal{O}(\mathrm{TeV})$. Thus the
hierarchies in neutrino masses, SM gauge boson masses and the
right-handed gauge boson masses require that
\begin{align}
v_L \ll \kappa \ll v_R \ .
\end{align}
An important consequnce of the smallness of $v_L$ is that the
couplings between left-handed triplets and SM gauge bosons $W^{\pm}$
and  $Z^0$  are strongly suppressed, as $\delta_L^{\pm\pm}W^{\mp}
W^{\mp}$, $\delta_L^{\pm}W^{\mp} Z^0$ and $\delta_L^{0}Z^0 Z^0$
couplings are all proportional to $v_L$.  For instance, the ratio 
between the decay width of $\delta_L^{++}\to \ell^+\ell^+$ and  $\delta_L^{++}\to
W^+W^+$ is given by
\begin{align}
\frac{\Gamma(\delta_L^{++}\to W^+W^+)}{ \Gamma(\delta_L^{++}\to
\ell^+\ell^+)} \approx \frac{g^4}{16} \left( \frac{v_L
m_{\delta_L}}{Y_{\ell\ell} m_W^2} \right)^2 \;,
\end{align}
where $Y_{\ell\ell}$ is the Yukawa coupling between $\delta_L^{++}$
and charged lepton $\ell^+$.  For the typical case $v_L\sim
\mathcal{O}(10^{-9})$ GeV and $m_{\delta_L}\sim \mathcal{O}(10^3)$
GeV, $\Gamma(\delta_L^{++}\to \ell^+\ell^+)$ is always dominant over
$\Gamma(\delta_L^{++}\to W^+W^+)$ as long as the Yukawa coupling is
not too small, namely, $Y_{\ell\ell}\gtrsim10^{-10}$.  Similar results
are found for the $W^{\pm} Z^0$, $Z^0 Z^0$ and $h^0 h^0$ final states.
Thus a small $v_L$ required by the tiny neutrino masses naturally
suppresses the triplet non-leptonic decays channels, such that our
model can avoid the constraints from the PAMELA anti-proton data.

\begin{figure}[htb]
\centerline{
\includegraphics[width=0.25\textwidth]{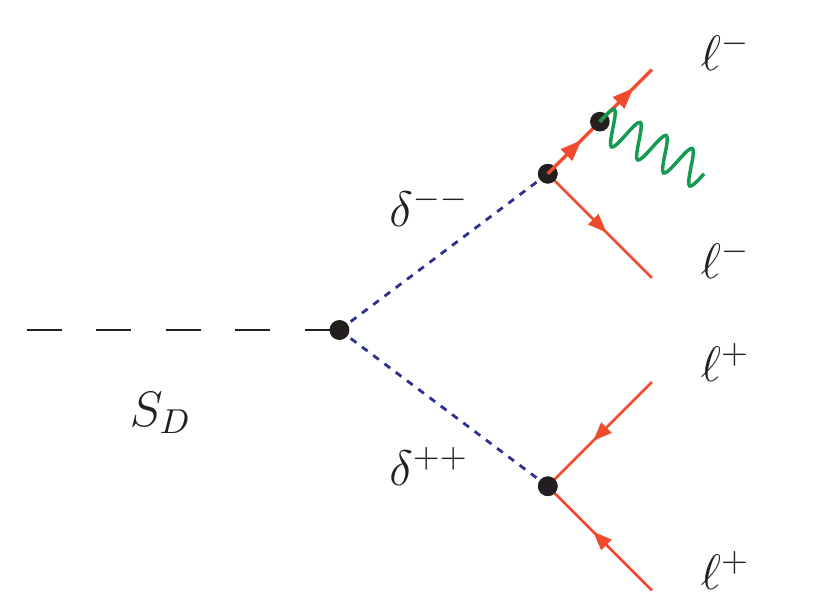}
\includegraphics[width=0.25\textwidth]{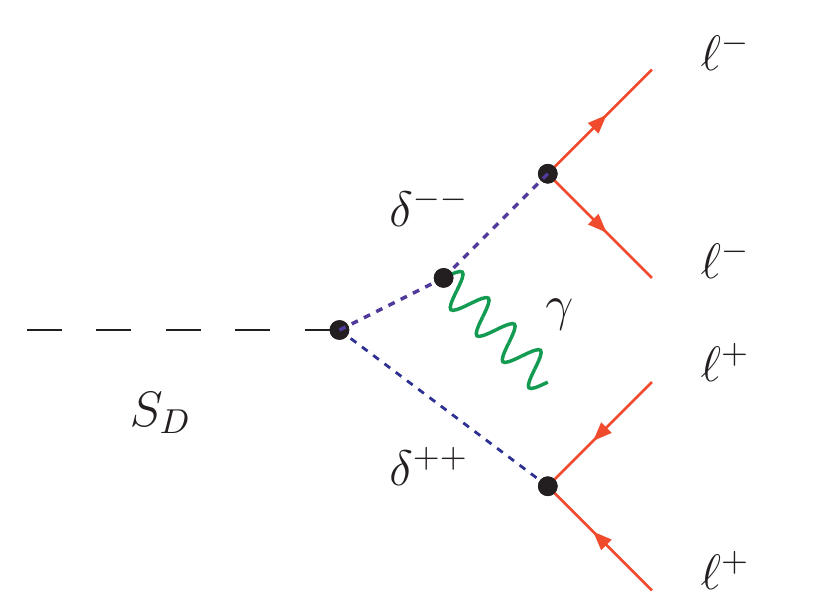}
} \caption{Feynman diagrams for gamma-ray emission through internal
bremsstrahlung. (Left)  final state radiation (FSR) from final state
charged leptons. (Right) virtual internal bremsstrahlung (VIB) from
doubly charged triplet scalars.} \label{gamma-ray}
\end{figure}

The Yukawa interaction term $\overline{\ell^{\mbox{ }c}_{L}} i\tau_2 \Delta_L
\ell_{L}$ requires that the couplings
$\delta^{\pm\pm}_{L}l^{\mp}l^{\mp}$, $\delta^{0}_{L}\nu\nu$ and
$\delta^{\pm}_{L}l^{\mp}\nu$ are the same within each generation in the flavor basis.
If the positron  excesses  reported  by the PAMELA and
Fermi-LAT are indeed from the DM  decay through triplets, there must exist associating high energy
neutrino flux which can be detected by the future experiments.
%

The $SU(2)$ triplet contains doubly as well as singly charged
scalars. Besides the ordinary ICS and FSR caused by the final state
charged leptons, the charged triplet scalars will emit gamma-rays
through VIB processes  in the cascade decay $S_D\to
\delta_L^{\pm\pm}\delta^{\mp\mp}\gamma$ as shown in Fig.
\ref{gamma-ray}. The photon multiplicity for a decay process $S\to
XX$ is defined as \cite{Bringmann:2007nk}
\begin{align}
\frac{dN_{XX}}{dx} \equiv \frac{1}{\sigma_{S\to
XX}}\frac{d\sigma_{S\to XX\gamma}}{dx} \;,
\end{align}
where $x=2E_\gamma/\sqrt{s}$ and $\sqrt{s}=m_S$ is the center-of-mass energy.
The radiation is dominated by the collinear  photon emission case, which can
be approximated by
\begin{align}
\frac{dN_{XX}}{dx}\approx \frac{\alpha Q^2}{\pi}
F(x)\log\left(\frac{s(1-x)}{m_X^2}\right) \label{VIB}
\end{align}
with $F(x)=(1+(1-x)^2)/x$ for fermions and $(1-x)/x$ for bosons. For
the decay process  $S_D\to \delta_L^{\pm\pm}\delta_L^{\mp\mp}$, one
has $s=m_{D}^2$ and $m_X=m_{\delta_L}$. For the sequential decay
$\delta_L^{\pm\pm}\to \ell^{\pm}\ell^{\pm}$, $s$ and $m_X$ are
replaced by $m_{\delta_L}$ and $m_\ell$ respectively.
For doubly charged particles there is a factor of four enhancement
relative to that of singly charged ones. The final positron and
gamma-ray energy spectra of cascade decays, such as $S_D\to
\delta_L^{\pm\pm}\delta_L^{\mp\mp}\to 4\ell$, depend on the mass of
the intermediate states, which also provide a possibility to probe
the masses of the triplets.


\section{Energy spectra of the cosmic-ray particles
\label{Sec:Propagation}}

The dark matter cascade decay processes $S_D\to \delta_L\delta_L\to
4\ell$, $2\ell 2\nu$ and $4 \nu_\ell$ contribute to new sources of
primary positrons, neutrinos and gamma-rays in our galaxy. Besides the
mass and lifetime of $S_D$, the final energy spectra of the cosmic-ray
particles depend on the mass ratio between $S_D$ and the triplets.  In
this section, we consider two typical mass hierarchies: the small mass
hierarchy case {\bf (SH)}, in which we take the mass ratio $r_m
\equiv2 m_{\delta_L}/m_{D}=0.8$, and the large hierarchy case {\bf
  (LH)} in which $r_m=0.1$.  Note that in the two extreme cases with
$r_m=$1 and $0$, the shape of the final energy spectra should reduce
to that of the two-body and four-body decays respectively. The
relative strengths of the Yukawa couplings also play important roles,
as they determine the corresponding branching ratios of different
decay modes.  In the basis in which the left-handed weak gauge
interaction is diagonal, the leptonic Yukawa couplings $Y_{ij}$
$(i,j=e,\mu,\tau)$ are related to $y_{ij}$ by $Y=U^T\cdot y\cdot U$,
where $U$ is the mixing matrix for charged lepton.  For the Yukawa
couplings, we are going to consider four representative cases. In each
of the first three cases, the Yukawa couplings are assumed to be
dominant by one of the three generation leptons, while in the last
case the couplings are assumed to be the same for all the generations,
i.e.,
$\mathbf{I}$) $|Y_{ee}|\gg |Y_{\mu\mu}|,|Y_{\tau\tau}|$, the triplet
scalars decay mainly into $4e$, $2e2\nu_e$ and $4\nu_e$;
$\mathbf{II}$) $|Y_{\mu\mu}|\gg |Y_{ee}|,|Y_{\tau\tau}|$, the
triplets decay mainly into $4\mu$, $2\mu2\nu_\mu$ and
$4\nu_\mu$;
$\mathbf{III}$) $|Y_{\tau\tau}|\gg |Y_{ee}|,|Y_{\mu\mu}|$, the
triplets decay mainly into $4\tau$, $2\tau2\nu_\tau$ and $4\nu_\tau$;
and $\mathbf{IV}$) $|Y_{ee}|\approx |Y_{\mu\mu}| \approx
|Y_{\tau\tau}| $, the triplet scalars decay to all leptons with the
same branching ratio. In fact, the case IV will give the same results
for the flavor democratic case in which all the Yukawa matrix elements
are nearly identical.  For simplicity we have assumed that the
off-diagonal Yukawa couplings $Y_{ij}\ (i \neq j)$ are all small such
that no lepton flavor violating process are considered.  For each
case, we consider various values of the mass and lifetime of $S_D$
with paying attention to the case that can reproduce the current data.
Note that in this model, the large mass hierarchy $r_m=0.1$ is
unlikely for the cases LH-I and LH-II unless $S_D$ is much heavier
than 10 TeV, since the lower bounds of the triplet masses are expected
to be around TeV scale. However, we still keep them for the sake of
completeness as such light intermediate states may be possible in
other models.

\subsection{Electrons and positrons }

In the Milky Way, the propagation of cosmic-ray (CR) particles can
be approximated by a two-zone propagation model, which includes two
cylinders centered at the galactic center. A cylinder of radius $R =
20$ kpc with height $2L$ ($L = 1-20$ kpc) in the $z$ direction
delimits the CR propagation region. A smaller cylinder with the same
radius but with thickness $2 h_z = 0.2$ kpc models the galactic
plane. The solar system is located at $r_\odot = 8.5$ kpc and
$z_\odot = 0$.  In general, the CR propagation equations contain
terms for convection and reacceleration effects
\cite{Strong:1998pw}. However, for electrons and positrons with
energy $E \gtrsim 10$ GeV, these effects can safely be neglected
\cite{Delahaye:2008ua}. Thus one only needs to consider the
diffusion effect. In this case, the steady-state diffusion equation
for positron is given by
\begin{eqnarray}
- K(E)\cdot \nabla^2 f_{e^+} - \frac{\partial}{\partial E} (b(E)
f_{e^+}) = Q \;, \label{SE}
\end{eqnarray}
where $f_{e^+} (E, r, z )$ is the positron differential number density in cylinder coordinate
and $K(E) = K_0 (E/{\rm GeV})^\delta$ is the spatial diffusion
coefficient. Through fitting the measured ratio of boron to carbon
(B/C), the propagation parameters $\delta$, $K_0$ and $L$ can be
determined \cite{Maurin:2001sj}. In Table. \ref{Propagation}, we list
three typical combinations of the propagation parameters
\cite{Delahaye:2008ua, Delahaye:2007fr}.  The parameter set MIN and
MAX corresponds to the minimal and maximal positron flux respectively,
while the  MED scenario best fits the B/C ratio and produce the
moderate  flux.   The energy loss rate $b(E)$
are mainly due to the ICS and synchrotron processes.
The two processes combined give $b(E) = E^2/({\rm GeV}\cdot \tau_E)$
with $\tau_E = 10^{16}$ s \cite{Baltz:1998xv}. For  decaying DM
model, the source term $Q$ is given by
\begin{eqnarray}
Q(r, E) =  \frac{\rho(r)}{m_{D}}  \sum_k \Gamma_k \frac{d
n_{e^+}^k}{d E} \;,
\end{eqnarray}
where $\Gamma_k$ is the decay width and $k$ runs over all the
channels with positrons in the final states. $d n_{e^+}^k/d E$ is
the positron energy spectrum per DM decay. The DM halo profile
$\rho(r)$ is usually parameterized as a spherically symmetric form
\begin{eqnarray}
\rho(r) = \rho_\odot \left( \frac{r_\odot}{r}\right)^\gamma \left(
\frac{1+ (r_\odot/r_s)^\alpha}{1+ (r/r_s)^\alpha}
\right)^{(\beta-\gamma)/\alpha}
\end{eqnarray}
with local DM density $\rho_\odot = 0.3 \, {\rm GeV cm}^{-3}$. Possible DM halo profile
parameters $\alpha, \beta, \gamma$ and $r_s$ are listed in
Table. \ref{Profile}. In this paper, we  adopt the NFW profile
and MED propagation model in numerical calculations.

\begin{table}
\begin{center}
\begin{tabular}{l  c c c }
 \hline \hline
Propagation & $\delta$  & $K_0$  ({\rm kpc$^2$/Myr})  & $L$ ({\rm kpc})  \\ \hline
 MED  & 0.70  & 0.0112 &  4  \\ 
 MAX  & 0.46  & 0.0765 & 15  \\ 
 MIN & 0.55  & 0.00595 & 1   \\ 
 \hline\hline
\end{tabular}
\end{center}
\caption{Three typical combinations for the propagation parameters
$\delta$, $K_0$ and $L$ from  the measured B/C ratio \cite{Delahaye:2008ua, Delahaye:2007fr}.
} \label{Propagation}
\end{table}

\begin{table}
\begin{center}
\begin{tabular}{l  l l l c}
 \hline \hline
Halo Profile \; & $\alpha$ \;\;\;\; & $\beta$ \;\;\;\; & $\gamma$ \;\;\;\;  & $r_s$ ({\rm kpc}) \\ \hline
 NFW  & 1  & 3 & 1 & 20 \\ 
 Isothermal  & 2  & 2 & 0 & 5 \\ 
 Moore  & 1.5  & 3 & 1.3 & 30 \\ 
 \hline\hline
\end{tabular}
\end{center} \caption{The values of parameters $\alpha, \beta, \gamma$ and $r_s$ for the NFW \cite{Navarro:1996gj},
Isothermal \cite{Bahcall:1980fb} and Moore \cite{Diemand:2004wh} DM
halo profiles. }\label{Profile}
\end{table}

Eq. (\ref{SE}) can be solved analytically from the above considerations.  For the decaying dark
matter, the solution of the positron diffusion  can be
written as \cite{Delahaye:2007fr, Hisano:2005ec}
\begin{eqnarray}
f_{e^+} (E, r, z) = \frac{1}{m_D} \int_E^{m_D} d E' G_{e^+} (E, E',
r, z) \sum_{k} \Gamma_k \frac{d n_{e^+}^k}{d E'} \;, \label{dnde}
\end{eqnarray}
The explicit form of the Green's function $G_{e^+}(E, E', r, z)$ is given
by
\begin{eqnarray}
G_{e^+}(E, E', r, z) & =& \sum_{n,m=1}^\infty  G_{nm}
\frac{\tau_E}{E^2} J_0 \left( \zeta_n \frac{r}{R} \right) \sin
\left[ \frac{m \pi}{2L} (z-L) \right] \\ \nonumber && \times {\rm
exp} \left[K_0 \tau_E \left(\frac{\zeta_n^2}{R^2} + \frac{m^2
\pi^2}{4 L^2} \right) \left(\frac{E^{\delta-1} - E'^{\delta
-1}}{\delta-1} \right) \right] \;,
\end{eqnarray}
with
\begin{eqnarray}
G_{nm} = \frac{2}{J_1^2(\zeta_n) R^2 L} \int_0^{R}  d r' r' J_0
\left( \zeta_n \frac{r'}{R} \right) \int_{-L}^L d z' \, \rho( r',
z')  \sin \left[ \frac{m \pi}{2L} (z'-L) \right] \;,
\end{eqnarray}
where $J_{0(1)}$ is the zeroth(first)-order  Bessel
function. $\zeta_n$ are successive zeros of the function $J_0$.  At
the heliosphere boundary ($r = r_\odot$ and $z=0$), the interstellar
positron flux per unit energy from the DM decay is
\begin{eqnarray}
\Phi_{e^+}^{DM} (E) = \frac{c}{4 \pi} f_{e^+} (E, r_\odot, 0 ) \;.
\end{eqnarray}

The interstellar $e^-$ and $e^+$ background is approximated by
\cite{Baltz:1998xv}
\begin{eqnarray}
\Phi_{e^-}^{prim}(E) &=& \frac{0.16 E^{-1.1}}{1+11E^{0.9} + 3.2
E^{2.15}} \;, \nonumber\\
\Phi_{e^-}^{sec}(E) &=& \frac{0.7 E^{0.7}}{1+ 110 E^{1.5} +
600E^{2.9} + 580 E^{4.2}}\;, \nonumber\\
\Phi_{e^+}^{sec}(E) &=& \frac{4.5 E^{0.7}}{1+ 650 E^{2.3} + 1500
E^{4.2}}\;,
\end{eqnarray}
where $E$ and $\Phi$ are in units of GeV and ${\rm GeV}^{-1} {\rm
  cm}^{-2} {\rm s}^{-1} {\rm sr}^{-1}$, respectively. The interstellar
(IS) positron fraction is given by
\begin{eqnarray}
PF (E_{IS}) = \frac{\Phi_{e^+}^{DM} + k_+
\Phi_{e^+}^{sec}}{\Phi_{e^+}^{DM} + \Phi_{e^-}^{DM} + k_+
\Phi_{e^+}^{sec} + k_- (\Phi_{e^-}^{prim} + \Phi_{e^-}^{sec})}
\;.
\end{eqnarray}
where we have introduced two real parameters $k_+$ and $k_-$ to
normalize the positron and electron background, which reflects the
background uncertainties \cite{Cirelli:2008pk}. In the numerical calculations we take
$k_{+}(k_{-})=0.9(0.7)$.

For the  solar modulation effects we adopt the Gleeson and Axford analytical force-field
approximation in which the final flux $\Phi (E_\oplus)$ at
the top of atmosphere is given by
\begin{eqnarray}
\Phi (E_\oplus) = \frac{p_\oplus^2}{p_{IS}^2} \Phi (E_{IS}) \;,
\end{eqnarray}
where  $\Phi (E_{IS})$ denotes the primordial interstellar flux.
$E_{IS}$ and $E_\oplus$ have the following relation
\begin{eqnarray}
E_{IS} = E_\oplus + |Z e| \phi_F  \;,
\end{eqnarray}
where $|Z|=1$ is the $e^\pm$ charge number. For the solar modulation
parameter $\phi_F$, we take a typical value of $\phi_F = 500$ MV.
We assume that the solar modulation effect is the same for electron
and positron, which leads to $PF (E_\oplus) = PF (E_{IS}) $.  Note
that even in the case of charge-independent modulation, the solar
modulation effect leads to a translation for the positron factor
$PF$, thus must be considered.

\begin{table}
\begin{center}
\begin{tabular}{l  c c l }
 \hline \hline
case & $m_{D}$ (TeV)  & $\tau_{D}$ ($10^{26}$ s)  &
$2m_{\delta_L}/m_{D}$\\ \hline
SH(LH)-{\rm I}   & 2.0     & 1.5  & 0.8 (0.1)\\
SH(LH)-{\rm II}  & 4.0     & 0.9  & 0.8 (0.1)\\
SH(LH)-{\rm III} & 8.0     & 0.4 & 0.8 (0.1)\\
SH(LH)-{\rm IV}  & 2.5     & 1.3  & 0.8 (0.1)\\
 \hline\hline
\end{tabular}
\end{center}
\caption{Typical values of the mass and lifetime of $S_D$ as well as
$m_{\delta_L}$ favored  by the PAMELA and Fermi-LAT (or ATIC) data
in different cases from SH(LH)-I to SH(LH)-IV.} \label{tab:param}
\end{table}

\begin{figure}[htb]\begin{center}
\includegraphics[scale=0.45]{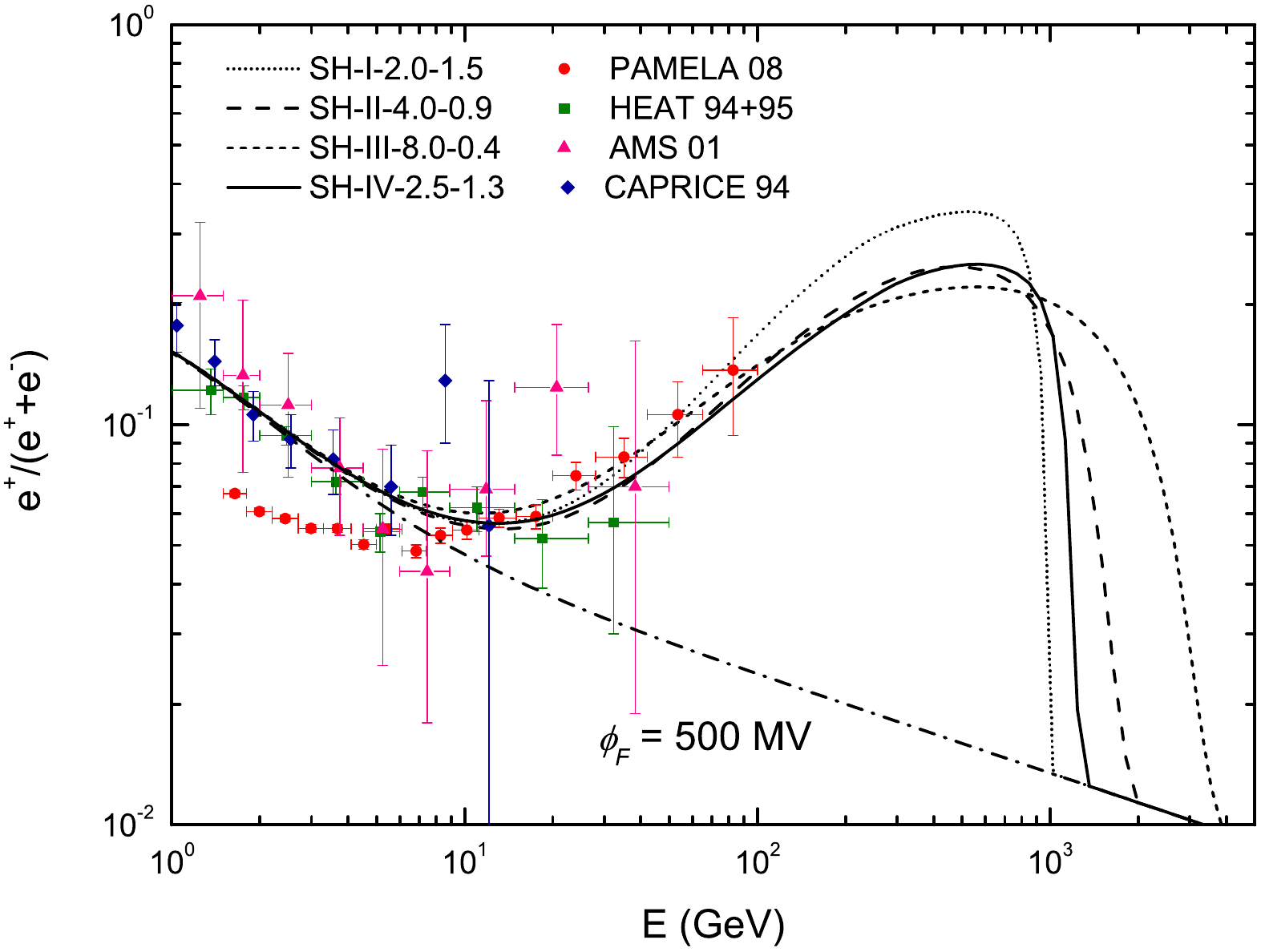}\includegraphics[scale=0.45]{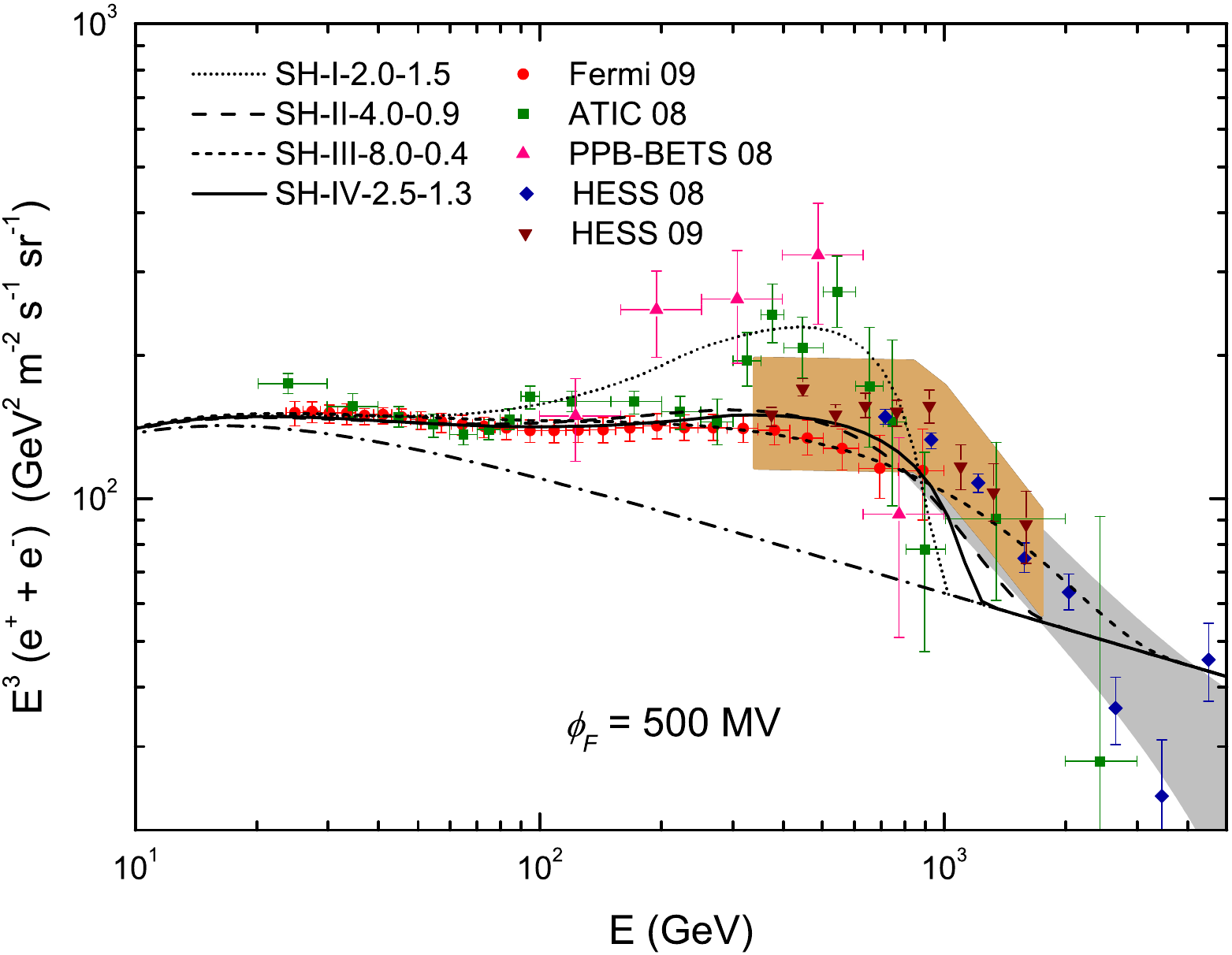}
\end{center}
\vspace{-0.5cm} \caption{The predicted positron fraction (left) and
the total electron and positron flux (right) for four representative
cases from SH-I to SH-IV. The relevant parameters are given in the figures in  compact notations.
For instance, the notation "SH-II-4.0-0.9" corresponds to the case of SH-II with $m_D$=4.0 TeV and
$\tau_D=0.9\times 10^{26}$s. }
\label{Fig:electron}\end{figure}

Making use of the above mentioned formulas we calculate the positron
fraction and the total electron and positron flux for the eight
cases SH(LH)-I$\sim$IV. The injection spectra $dn^k_e/dE$ is
obtained by using the package Pythia 8.1 \cite{Sjostrand:2007gs}. In
each case we consider some typical values of the parameters $m_{D}$
and $\tau_{D}$ which can reproduce the experimental data well. The
favored values for the parameters are listed in Tab. \ref{tab:param}
and the corresponding numerical results are given in Fig.
\ref{Fig:electron}. From the case SH(LH)-I to SH(LH)-III, the
required mass of $S_D$ increases from 2 to 8 TeV while the lifetime
decreases from $\tau_D=1.5\times 10^{26}$ s to $\tau_D=0.4\times 10^{26}$ s. The
case SH(LH) -IV is a kind of mixture with equal Yukawa couplings for
all generation leptons, and the required mass and lifetime are found
close to the case SH(LH)-I. As indicated in Fig. \ref{Fig:electron},
the SH-I case can explain both the PAMELA and ATIC data. In this
case, the DM particle decays directly into  $4e$, which usually leads
to a too hard spectrum to meet the Fermi-LAT data. The SH-II$\sim$IV
cases involving $\mu$- and $\tau$-lepton as final states can
reproduce the PAMELA and Fermi-LAT results well with the parameters
in Tab. \ref{tab:param}.  The differences among the four cases are
not obvious in the low energy rang $E \lesssim 100$ GeV. However, in
higher energy range $ E \gtrsim 1$ TeV, the heavier final state case
leads to harder energy spectrum. The SH-III case with $4\tau$ final
states has the hardest spectrum among all the cases in this energy
region.

\begin{figure}[htb]\begin{center}
\includegraphics[scale=0.8]{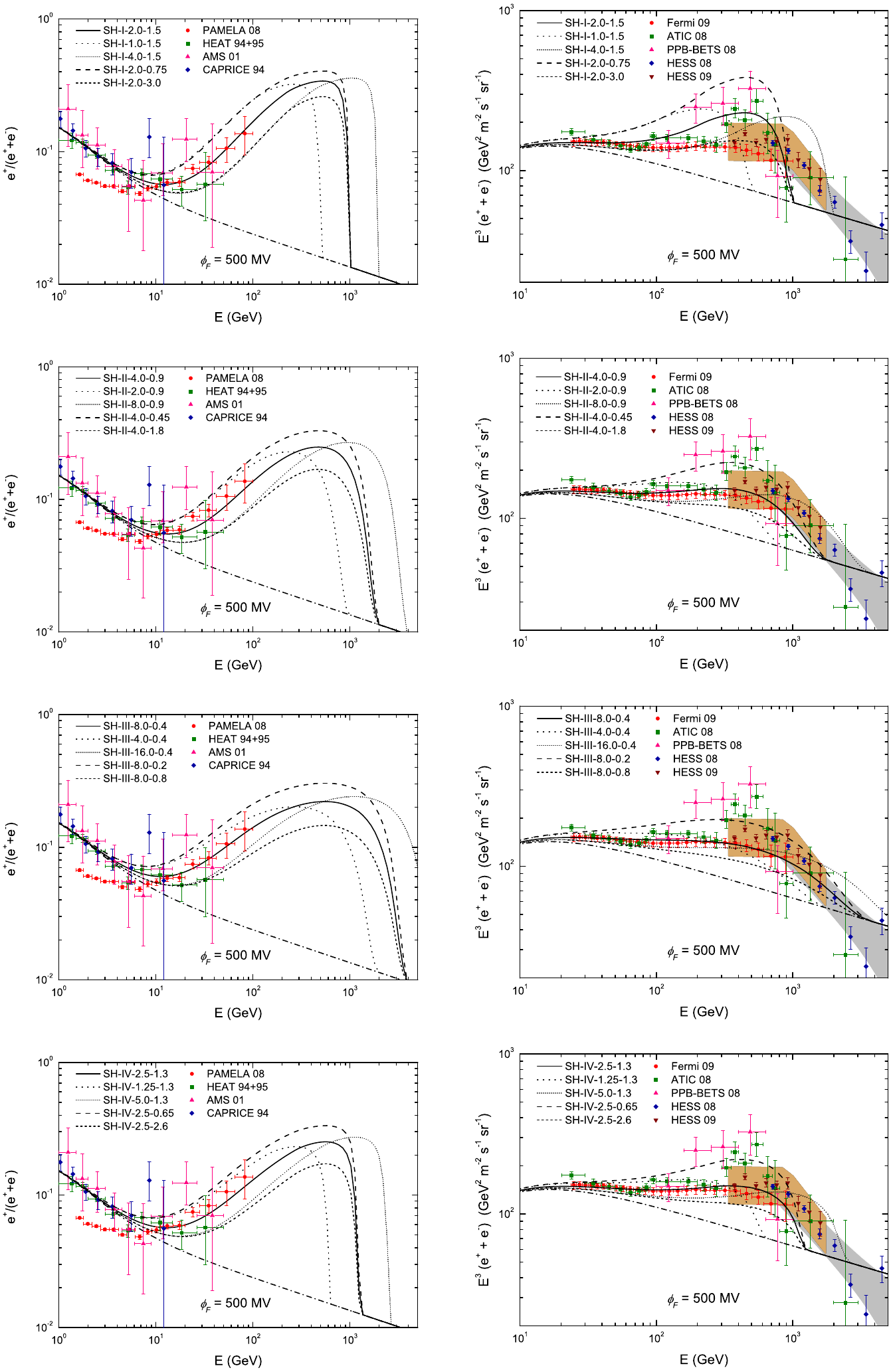}
\end{center}
\vspace{-0.5cm} \caption{Effects of the DM particle mass and
lifetime  on the positron fraction (left) and the total
electron and positron flux (right) for four representative cases from SH-I to SH-IV. For each case
the mass and life-time of $S_D$ are varied by a factor of two.}
\label{Electron1}\end{figure}

\begin{figure}[htb]\begin{center}
\includegraphics[scale=0.8]{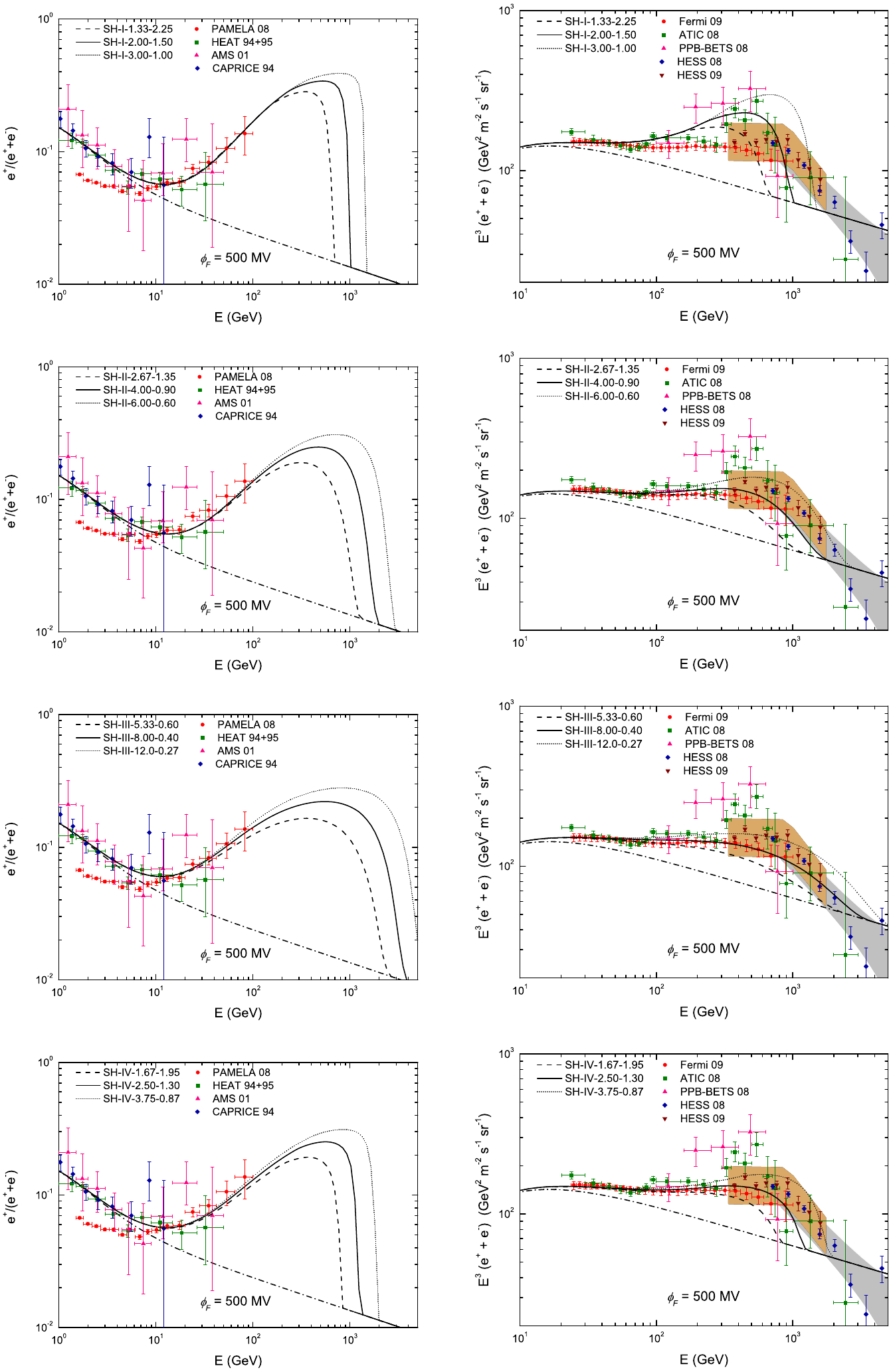}
\end{center}
\vspace{-0.5cm} \caption{The same as Fig. \ref{Electron1}, but
for  different  mass and lifetime of $S_D$ which can have nearly the same results for 
the positron fraction but significantly differ in the total flux of electron and positron. See text
for detailed explanation.  }
\label{Electron2}\end{figure}

The mass and lifetime of $S_D$ are two key parameters in determining
the positron fraction and the total electron and positron flux. In
Fig.  \ref{Electron1}, we vary the DM mass or lifetime by a factor of
two for each case listed in Tab. \ref{tab:param}. The results show
that in general the location of the peak in the spectrum is controlled
by $m_{D}$, and the height of the peak is more relevant to $\tau_{D}$.
Compared with the PAMELA data, the Fermi-LAT data is more sensitive to
$m_{D}$, which is due to more data points at high energy region and
higher statistics. If simultaneously adjusting both $m_D$ and
$\tau_D$, in all the cases from SH-I to SH-IV, one can obtain the
nearly the same positron fraction ($E \lesssim 100$ GeV) to explain
the PAMELA data as shown in Fig.  \ref{Electron2}. However, the
predicted total electron and positron flux are obvious different.

\begin{figure}[t]\begin{center}
\includegraphics[scale=0.45]{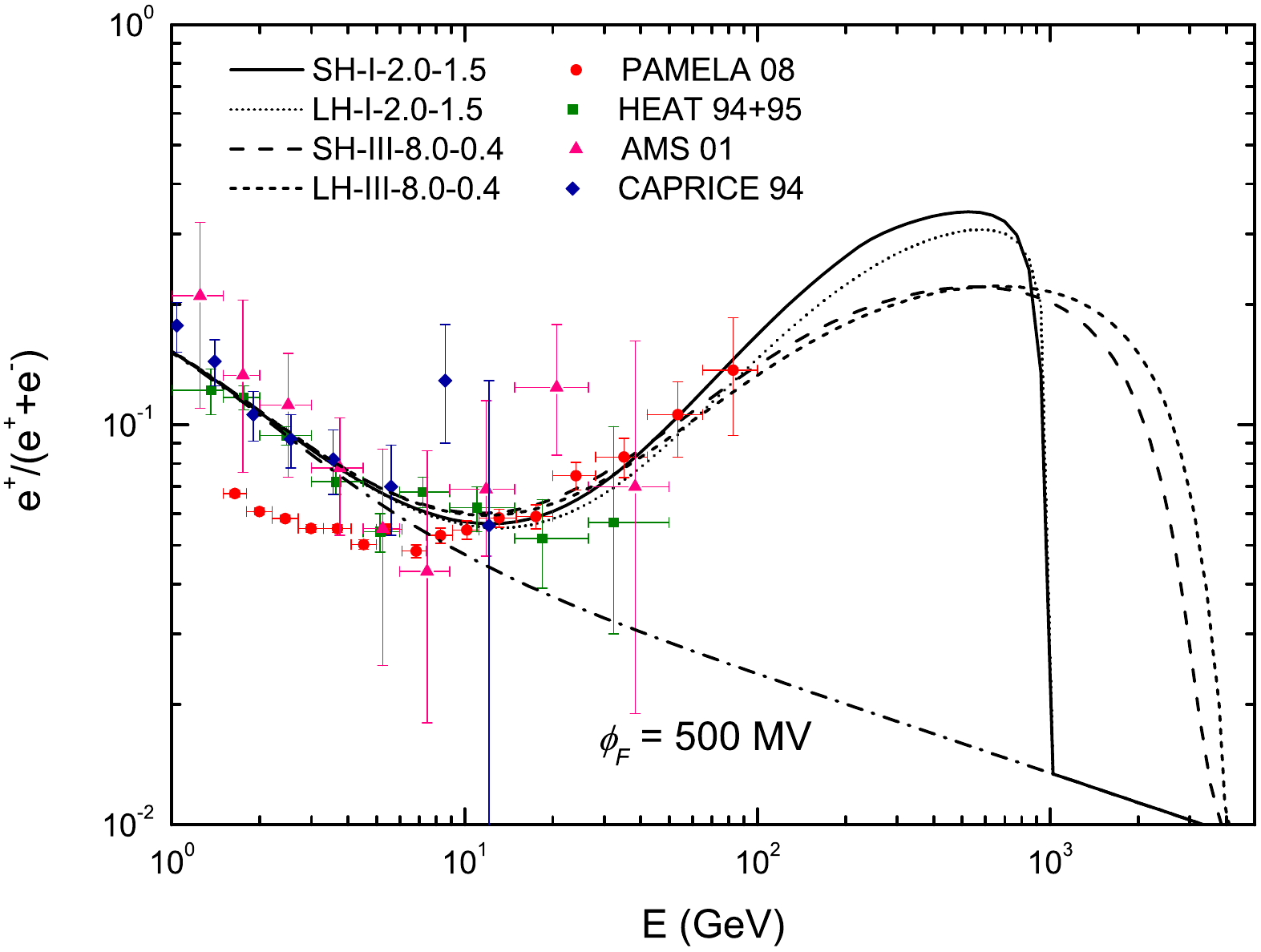}\includegraphics[scale=0.45]{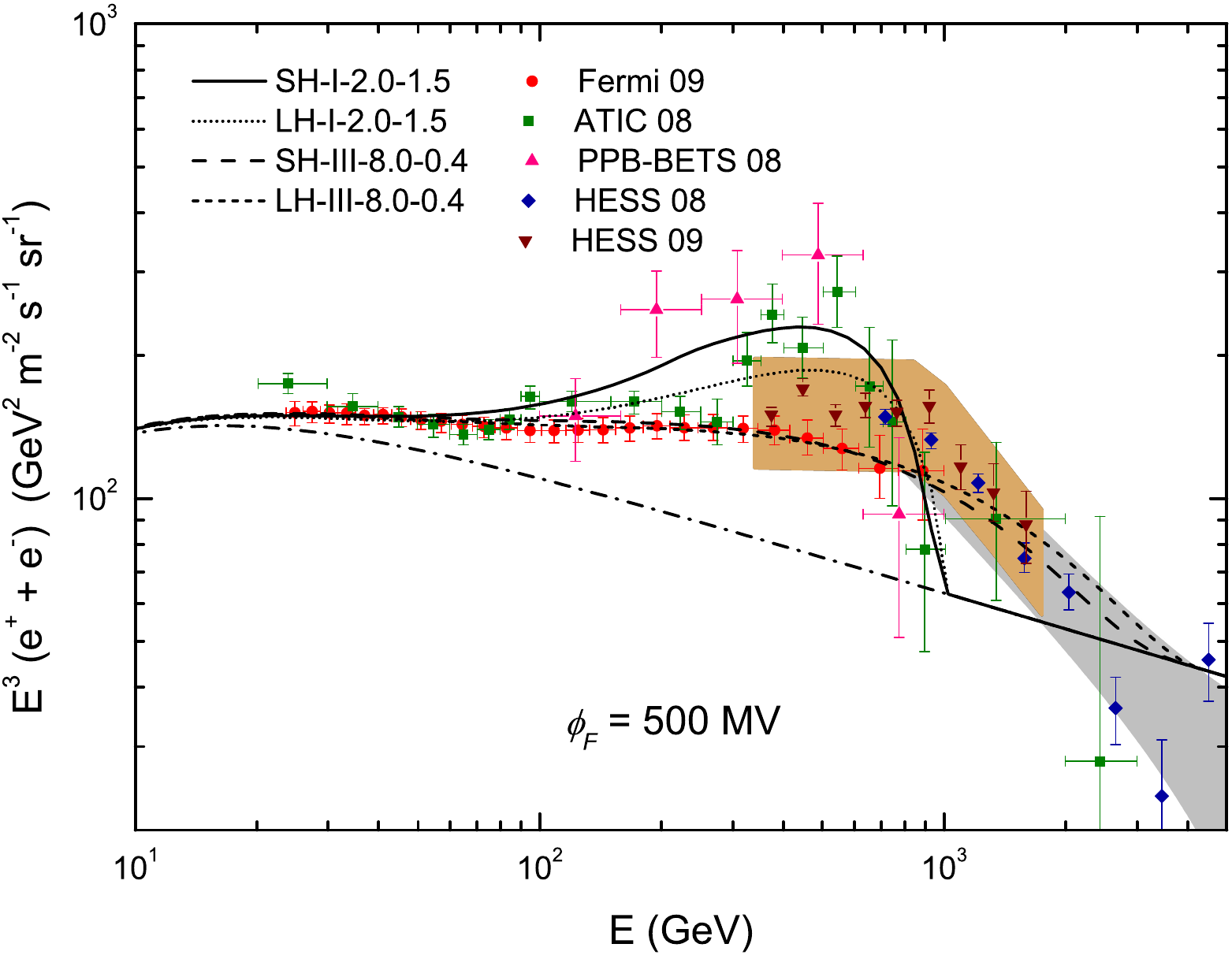}
\end{center}
\vspace{-0.5cm} \caption{Difference 
between the small and large hierarchy scenarios in the positron fraction (left)
and the total flux of electron and positron (right)  for the cases SH(LH)-I and SH(LH)-II.  }
\label{FigESHLH}\end{figure}

For a comparison between SH and LH cases, we show in Fig.
\ref{FigESHLH} the results for SH(LH)-I and SH(LH)-III. It is clear
that the differences are rather mild for the parameters given in Tab.
\ref{tab:param}. Actually, the SH and LH cases have significant
differences in the injection energy spectrum $d n_{e^+}^k/d E$.
However, the differences are smeared out due to the energy loss
processes from the ICS and the synchrotron.

\subsection{Neutrinos}

The expected muon neutrino flux coming from the galactic center (GC)
and arriving at the Earth can be estimated by \cite{Hisano:2008ah}
\begin{eqnarray}
\frac{d \Phi_{\nu_\mu}}{d E_{\nu_\mu}}  = \rho_\odot r_\odot
\frac{1}{4 \pi m_D} \left( \sum_{\alpha=e,\mu, \tau} P_{\nu_\alpha
\rightarrow \nu_\mu} \sum_k \Gamma_k \frac{d n_{\nu_\alpha}^k}{d
E_{\nu_\alpha}} \right ) J_{\Delta \Omega} \Delta \Omega \;,
\end{eqnarray}
where the neutrino oscillation probabilities $P_{\nu_\alpha
  \rightarrow \nu_\mu}$ are $P_{\nu_e \rightarrow \nu_\mu} = 0.22$,
$P_{\nu_\mu \rightarrow \nu_\mu} = 0.39$ and $P_{\nu_\tau \rightarrow
  \nu_\mu} = 0.39$ \cite{Strumia:2006db}. One can have the same
results for the muon anti-neutrino flux $\Phi_{\bar{\nu}_\mu}$. The
averaged number flux $J_{\Delta \Omega}$ in a cone with half-angle
$\phi$ around GC is given by
\begin{eqnarray}
J_{\Delta \Omega} = \frac{1}{\rho_\odot r_\odot} \frac{1}{\Delta
\Omega} \int_{\Delta \Omega} d \Omega \int_{LOS} \rho(l) d l \;,
\label{J}
\end{eqnarray}
where $\Delta \Omega = 2 \pi (1- \cos \phi)$ is the solid angle. The
above equation can be written as the following form
\begin{eqnarray}
J_{\Delta \Omega} = \frac{1}{\rho_\odot r_\odot} \frac{2\pi}{\Delta
\Omega} \int_{\cos \phi}^1  d \cos \phi'  \int_0^{l_{max}} \rho
\left(\sqrt{r_\odot^2 - 2 l r_\odot \cos \phi' + l^2} \right) d l
\;,
\end{eqnarray}
where the integration upper limit $l_{max} = \sqrt{r_{MW}^2 -
r_\odot^2 \sin^2 \phi'  } + r_\odot \cos \phi'$ and the DM halo size
$r_{MW} \approx 30$ kpc.

The muon neutrinos interact with the earth rock to produce the
upgoing muon flux, which can be detected by the water Cherenkov
detector Super-Kamiokande (SK) \cite{Desai:2004pq, Mardon:2009rc}.
The neutrino induced muon flux is give by \cite{Barger:2007xf}
\begin{eqnarray}
\Phi_\mu = \int_{E_{thr}}^{m_D/2} d E_{\nu_\mu} \frac{d
\Phi_{\nu_\mu}}{d E_{\nu_\mu}} \int_{E_{thr}}^{E_{\nu_\mu}} d E_\mu
\, L(E_\mu) \sum_{a=p,n} n_a \sum_{x=\nu_\mu, \bar{\nu}_\mu} \frac{d
\sigma_x^a (E_{\nu_\mu})}{d E_\mu} \;.
\end{eqnarray}
where $L(E_\mu)$ is the range of a muon with energy $E_\mu$ until its
energy drops below the SK threshold $E_{thr} = 1.6$ GeV:
\begin{eqnarray}
L(E_\mu) = \frac{1}{\rho \beta_\mu} \ln \frac{\alpha_\mu + \beta_\mu
E_\mu}{\alpha_\mu + \beta_\mu E_{thr}} \;,
\end{eqnarray}
where $\alpha_\mu = 2 \times 10^{-3} \, {\rm g}^{-1}\, {\rm GeV}\,
{\rm cm}^2 $ and $\beta_\mu = 4.2 \times 10^{-6} \, {\rm g}^{-1}
\,{\rm cm}^2$. $\rho$ is the density of the material in ${\rm g \,
cm}^{-3}$ and $n_a \approx r_a \rho/m_p $ are the number densities
of neutrons and protons with $r_p \approx 5/9$ and $r_n \approx
4/9$. For the detection cross section, we use
\begin{eqnarray}
\frac{d \sigma_x^a (E_{\nu_\mu})}{d E_\mu} \approx \frac{2 m_p
G_F^2}{\pi} \left(  A_x^a + B_x^a \frac{E_\mu^2}{E_{\nu_\mu}^2}
\right) \;,
\end{eqnarray}
where $A_{\nu_\mu}^{n,p} = 0.25, 0.15$, $B_{\nu_\mu}^{n,p} = 0.06,
0.04$ and $A_{\bar{\nu}_\mu}^{n,p} = B_{\nu_\mu}^{p,n}$,
$B_{\bar{\nu}_\mu}^{n,p} = A_{\nu_\mu}^{p,n}$.

\begin{figure}[t]\begin{center}
\includegraphics[scale=0.45]{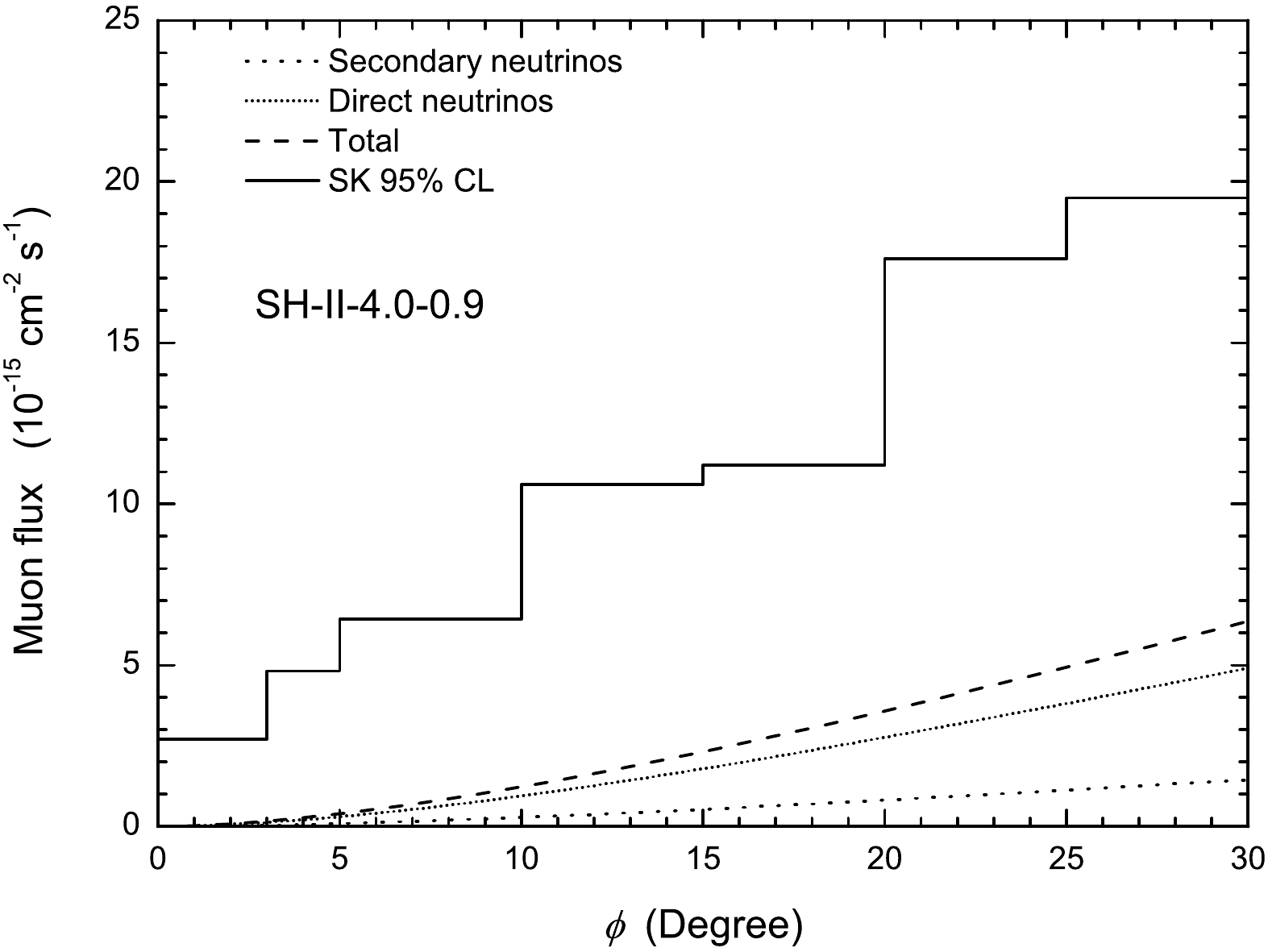}
\includegraphics[scale=0.45]{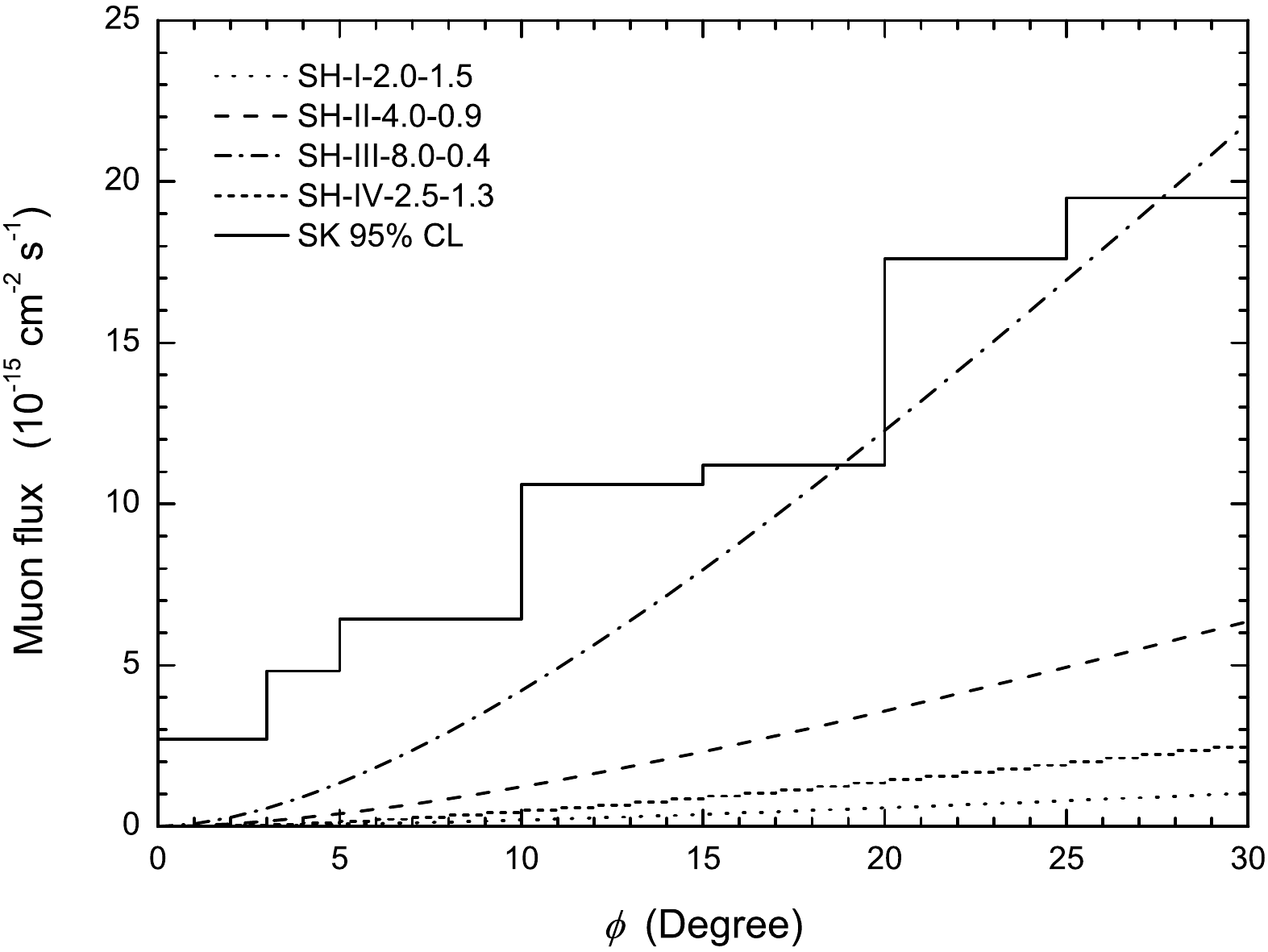}
\end{center}
\vspace{-0.5cm} \caption{The predicted neutrino-induced  upgoing
muon flux as a function of the cone half-angle around the galactic
center. (left)  the direct and
secondary neutrino contributions in the SH-II case.  (Right)
the predicted total neutrino-induced  upgoing
muon flux for the cases SH-I $\sim$ SH-IV.}
\label{FigNeutrino}\end{figure}

As mentioned in Sec II, an important feature of this model is that the
Yukawa couplings for $\delta_L^{++}\ell^{-}\ell^{-}$,
$\delta_L^+\ell^-\nu_\ell$ and $\delta_L^0 \nu_\ell\nu_\ell$ are
equal.  In the DM particle decay processes, the final state neutrinos
may come from the direct decay of $\delta_L$ and from the secondary
decay of charged leptons.  If a large decay rate of
$\delta_L^{++}\to\ell^{+}\ell^{+}$ is needed for explaining the
current experimental data, the associated processes
$\delta_L^+\to\ell^+\nu_\ell$ and $\delta_L^0 \to\nu_\ell\nu_\ell$
will give extra contributions to the neutrino flux, which can be more
important than that neutrinos from the secondary decay of charged
leptons.  In Fig. \ref{FigNeutrino}, we compare the contributions to
the total neutrino-induced muon flux from the direct neutrinos and
that from the secondary lepton decays in the SH-II case. The
difference can be clearly seen in the large angle region $\phi\simeq
30^\circ$ in which the  direct neutrino contribution is roughly five times as
many as that from secondary charged lepton decays. In
Fig. \ref{FigNeutrino},  we also give the predicted muon flux for the
cases SH-I $\sim$ SH-IV.  The results show that the SH-III case with
$S_D$ decaying into $4\tau,2\tau 2\nu_\tau$ and $4\nu_\tau$ predicts
much larger flux than the other cases. The reason is that the direct
neutrinos in this case have higher energy than that in other
cases. This case can be tested in the future experiments with improved
sensitivity by a factor of a few. Note that the $2\tau$ or $4\tau$
final state cases are now strongly disfavored by the latest Fermi-LAT
gamma-ray data \cite{newResults} even in the DM decay scenario. The
predictions for the cases LH-I $\sim$ LH-IV are found to be very
similar.

\subsection{Gamma-rays \label{SecGamma}}

The high energy gamma-rays generated by DM particle may come from two
type of contributions. The first one is that the DM particle directly
produces high energy gamma-rays through processes VIB and FSR. The
second contribution comes from the ICS process. In this case, the low
energy interstellar photons obtain energy from the high energy
positrons and electrons. For the first contribution, the expected
spectrum of high energy gamma-rays is given by
\begin{eqnarray}
\frac{d \Phi_{\gamma}}{d E_{\gamma}}  = \frac{\rho_\odot r_\odot}{4
\pi m_D}\;  J_{\Delta \Omega}  \left( \sum_k \Gamma_k \frac{d
n_{\gamma}^k}{d E_{\gamma}} \right )  \;,
\end{eqnarray}
where $J_{\Delta \Omega}$ is defined in the same way as for neutrino
in Eq. (\ref{J}).  For a general observable region,  it is
convenient to  choose the galactic coordinate $(b,l)$ where $b$ and
$l$ stand for the galactic latitude and longitude. Then $J_{\Delta
\Omega}$ can be written as the following form
\begin{eqnarray}
J_{\Delta \Omega} = \frac{1}{\rho_\odot r_\odot} \frac{1}{\Delta
\Omega} \int_{\Delta \Omega} d \Omega  \int_0^{s_{max}} \rho
\left(\sqrt{r_\odot^2 - 2 s r_\odot \cos b \cos l + s^2} \right) d
s\;,
\end{eqnarray}
where the integration upper limit $s_{max} = \sqrt{r_{MW}^2 -
r_\odot^2 (1 - \cos^2 b \cos^2 l ) } + r_\odot \cos b \cos l$ and
$d\Omega=\cos b db dl$.

The high energy gamma-rays produced by the ICS process $e^{\pm}
\gamma \rightarrow e^{\pm '} \gamma'$  have the following energy
spectrum \cite{Cirelli:2009vg,Meade:2009iu}:
\begin{eqnarray}
\frac{d \Phi_{\gamma'}}{d E_{\gamma'}} = \frac{\alpha_{em}^2}{\Delta
\Omega} \int_{\Delta \Omega} d \Omega \int_{LOS} d s \int \int
f_{e^+} (E_e, r, z ) \; u_{\gamma} (E_\gamma, r, z) \; f_{ICS}
\frac{d E_e}{E_e^2 } \frac{ d E_\gamma}{E_\gamma^2} \;.
\label{Gamma}
\end{eqnarray}
Since the photon may be emitted from both electron and positron,  an
overall factor of $2$ has been multiplied in the equation. The
differential energy density $u_{\gamma}(E_\gamma, r, z)$ of
interstellar radiation field (ISRF)  contains three components:
the cosmic microwave background (CMB), thermal dust radiation and
star light. Here we adopt  the GALPROP numerical result for $u_{\gamma} $ in Ref. \cite{Porter:2005qx}. The parameter
$f_{ICS}$ is defined by \cite{Cirelli:2009vg}
\begin{eqnarray}
f_{ICS} = 2 q \ln q + (1 + 2 q) (1-q) + \frac{1}{2}
\frac{\epsilon^2}{1-\epsilon} (1-q) \;,
\end{eqnarray}
with
\begin{eqnarray}
\epsilon = \frac{E_{\gamma'}}{E_e}, \;\;\;\;\; q = \frac{E_{\gamma'}
m_e^2}{4 E_{\gamma} E_e (E_e -E_{\gamma'})} \;,
\end{eqnarray}
where $0 \leq q \leq 1$. In Eq. (\ref{Gamma}), $f_{e^+} (E_e, r, z
)$ is the positron differential number density. Because of the ICS
and synchrotron, high energy electrons and positrons will loss most
of their energy within a kpc. Therefore one may neglect the
diffusion term of Eq. (\ref{SE}) and approximately calculate
$f_{e^+} (E_e, r, z )$  for every point in the CR propagation
region. In this case, $f_{e^+} (E_e, r, z )$ is given by the
following formulas \cite{Cirelli:2009vg,Ishiwata:2009dk}
\begin{eqnarray}
f_{e^+} (E_e, r, z) = \frac{1}{b_{ICS} (E_e, r, z)} \frac{\rho(r,
z)}{m_D} \sum_{k} \Gamma_k \int_{E_e}^{m_D} d E' \frac{d
n_{e^+}^k}{d E'} \;,
\end{eqnarray}
where the electron energy loss rate $b_{ICS} (E_e, r, z)$ is
\begin{eqnarray}
b_{ICS} (E_e, r, z) = \frac{2 \pi \alpha_{em}^2}{E_e^2} \int  d
E_{\gamma} \frac{u_{\gamma} (E_\gamma, r, z)}{E_{\gamma}^2} \int d
E_{\gamma'} (E_{\gamma'}-E_{\gamma}) f_{ICS}  \;.
\end{eqnarray}
Here we neglect  the synchrotron energy loss rate $b_{syn}$
as $b_{syn} \ll b_{ICS}$ \cite{Meade:2009iu}. It is worthwhile to
stress that the above energy loss rate $b_{ICS} (E_e, r, z)$ is
position dependent\footnote{If a position independent energy loss
rate is assumed, one can use Eq. (\ref{dnde}) to calculate $f_{e^+}
(E_e, r, z )$.}.


\begin{figure}[t]\begin{center}
\includegraphics[scale=0.80]{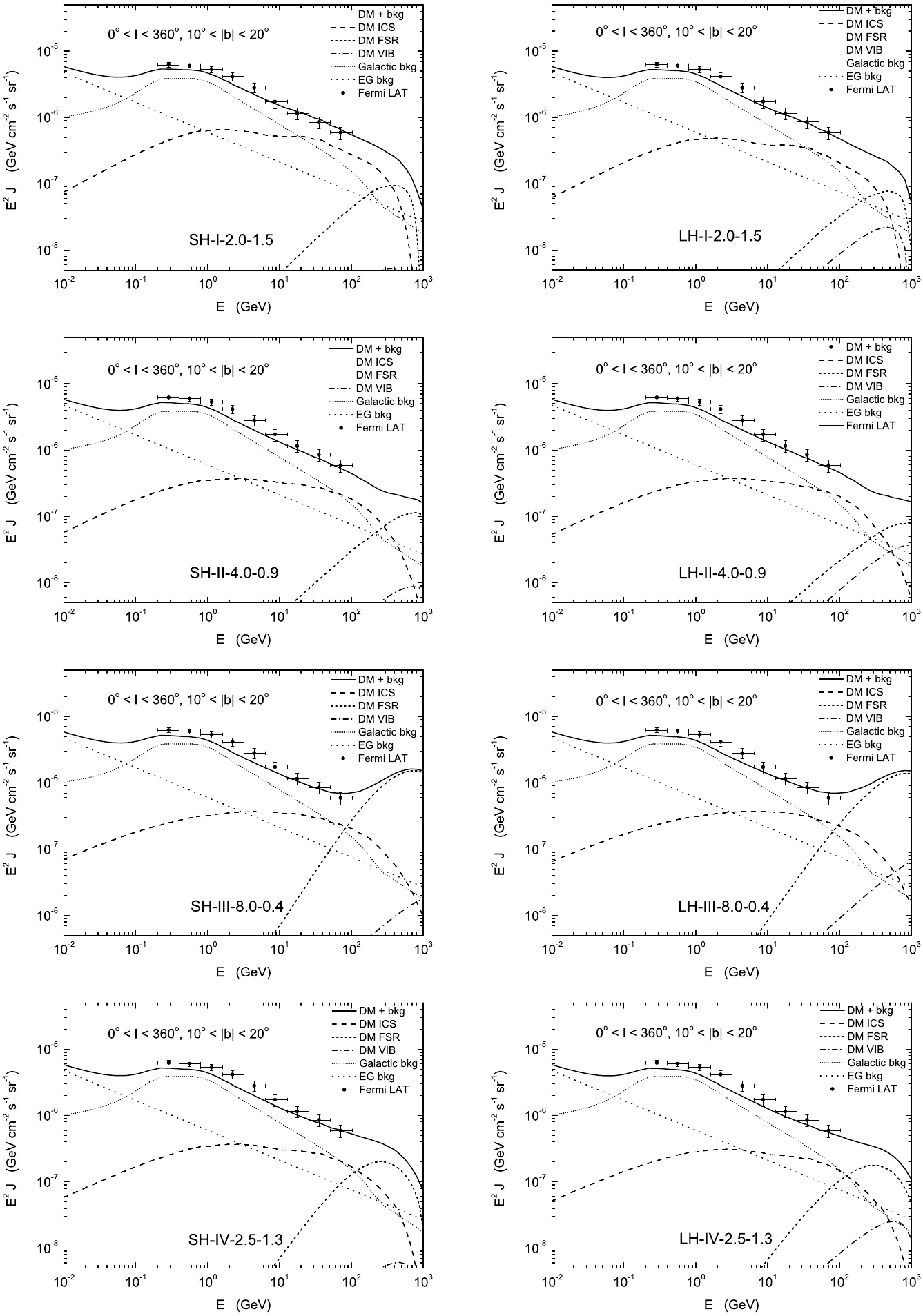}
\end{center}
\caption{The predicted gamma-ray spectra in  the small hierarchy
(left) and large hierarchy (right) scenarios for the
observed region $10^\circ < |b| <  20^\circ$. }
\label{FigGamma1020}\end{figure}

\begin{figure}[t]\begin{center}
\includegraphics[scale=0.80]{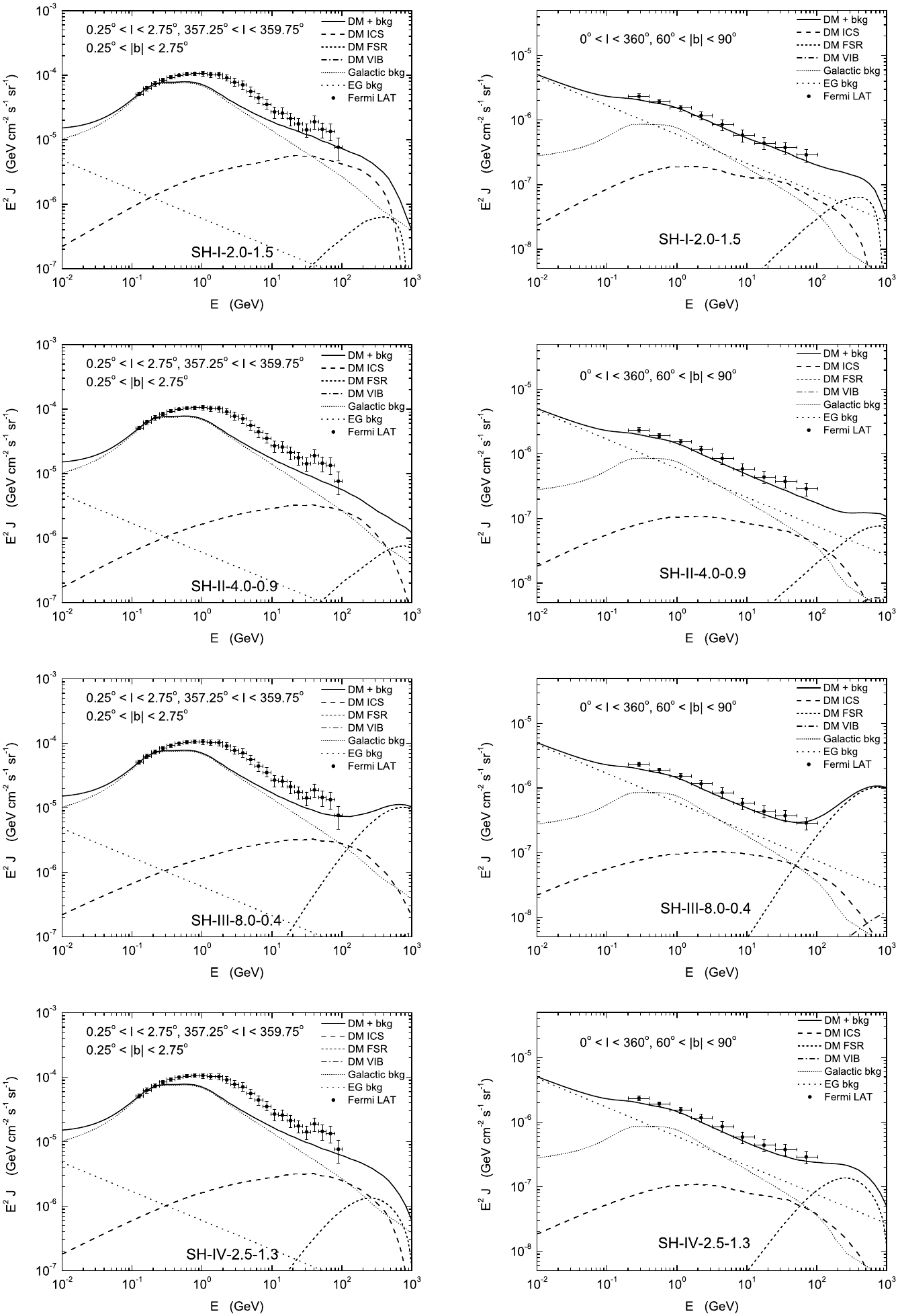}
\end{center}
\caption{The predicted gamma-ray spectra for the observed regions
$0.25^\circ < |b| < 2.75^\circ$, $0.25^\circ < |l| < 2.75^\circ$,
$357.25^\circ < |b| < 359.75^\circ$(left) and $60^\circ < |b|
< 90^\circ$,  $0^\circ < |l| < 360^\circ$ (right) in the small
hierarchy  scenario for the cases SH-I $\sim$ SH-IV.} \label{FigGamma2669}\end{figure}

In order to compare with the experimental data, one needs to know the
diffuse gamma-ray background which includes a galactic
$\Phi_\gamma^{\rm Galactic}$ contribution and an extragalactic (EG)
$\Phi_\gamma^{\rm EG}$ contribution. The galactic gamma-ray background
$\Phi_\gamma^{\rm Galactic}$ mainly comes from pion decay, ICS and
bremsstrahlung.  In principle, one should run the GALPROP code
\cite{Strong:1998pw} to obtain $\Phi_\gamma^{\rm Galactic}$ for given
diffusion parameters $\delta$, $K_0$ and $L$.  Note that
$\Phi_\gamma^{\rm Galactic}$ is not sensitive to these given diffusion
parameters as shown in Refs. \cite{Regis:2009md}.  For an
illustration, we use the numerical results of the conventional GALPROP
model (44\_500180) in Ref. \cite{Strong:2004de} as an estimate of our galactic
gamma-ray background $\Phi_\gamma^{\rm Galactic}$.  For the diffuse
extragalactic gamma-ray background, one can adopt the following
parametrization from the Fermi-LAT preliminary results
\cite{Ackermann}:
\begin{eqnarray}
\Phi_\gamma^{\rm EG} = 6.0 \times 10^{-7} \left
(\frac{E_\gamma}{{\rm GeV}}\right)^{-2.45} \;. 
\end{eqnarray}

In Fig. \ref{FigGamma1020}, we show the contributions from ICS, FSR
and VIB in various cases in the sky region  $10^\circ
<|b|<20^\circ$, $0^\circ<l<360^\circ$. For all the cases in lower
energy region $E\lesssim 100$ GeV, the DM contributions are
dominated  by the ICS process which is typically $1\sim2$ order of
magnitude below the current Fermi-LAT data \cite{Ackermann}.  The
ICS process contributes to a broad spectrum from $10^{-2}$ GeV to a
few hundred GeV.  At higher energy region $E\sim 10- 100$ GeV, the
ICS contribution becomes significant as the background drops
rapidly. The ICS has no significant dependence on the mass hierarchy
as it only depends on the spectrum of the final state electrons. In
general, compared with ICS processes, VIB and FSR contribute to
gamma-rays at higher energy. For photon energy above 500 GeV, they
become dominant sources and may lead to an up turn of the photon
spectrum. This prediction can be tested by the future gamma-ray
detectors. The similar conclusions can be obtained for the galactic
center region ($0.25^\circ <|b|<2.75^\circ$,
$0.25^\circ<l<2.75^\circ$, $357.25^\circ<l<359.75^\circ$) and the
galactic pole region ($60^\circ <|b|<90^\circ$,
$0^\circ<|l|<360^\circ$). As shown in Figs. \ref{FigGamma1020} and
\ref{FigGamma2669}, the DM particle decay can give significant contributions
to the high energy gamma-rays for the three typical regions. For
an illustration purpose, we naively sum up the DM contributions (ICS, FSR and
VIB) and the diffuse gamma-ray background ($\Phi_\gamma^{\rm
Galactic}$ and $\Phi_\gamma^{\rm EG}$). One finds that  all the
cases are compatible with the current Fermi-LAT preliminary results
\cite{Ackermann,Digel}.

\begin{figure}[htb]\begin{center}
\includegraphics[scale=0.5]{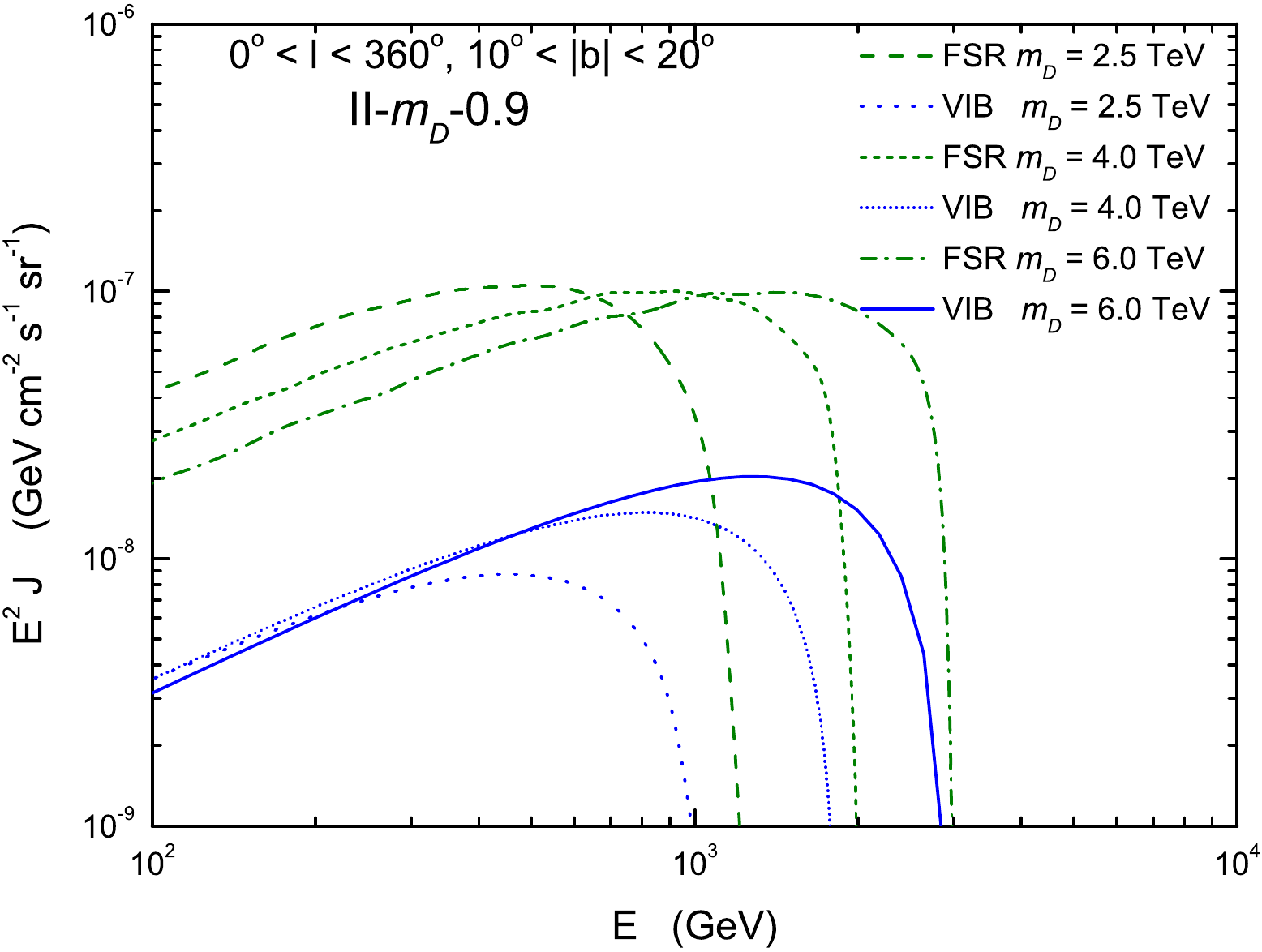}
\end{center}
\vspace{-0.5cm} \caption{Comparison of the predicted  energy spectra 
between FSR and VIB process. The curves corresponds to  $S_D=2.5, 4.0$ and $6.0$ TeV respectively, 
with the mass of triplet scalar $m_{\delta_L}$ fixed at 1 TeV and $\tau_D=0.9\times 10^{26}$s.  }
\label{Fig:VIBvsFSR}\end{figure}

The FSR and VIB processes show stronger dependence on the masses of
the triplet scalars, and the two contributions are correlated. It
follows from Eq. (\ref{VIB}) that for fixed masses of initial DM
particle and final states, the decrease of the triplet mass will
enhance VIB while slightly suppress FSR. Thus the VIB process can be
important for the case with large mass hierarchy between $S_D$ and
$\delta_L$. In Fig. \ref{Fig:VIBvsFSR}, we give the FSR and VIB
contributions to the gamma-ray spectrum for three different DM mass
cases ($m_{D}=2.5, 4$ and 6 TeV). Here we have fixed the triplet
mass $m_{\delta_L} = 1$ TeV and the DM lifetime $\tau_D = 0.9 \times
10^{26}$ s. We can see that with $m_{D}$ increasing, the VIB
contribution becomes more significant relative to FSR. For a low
$m_{D}=2.5$ TeV, the contribution from VIB relative to that of FSR
is  only $\sim 7\%$. But for a larger $m_{D}=6$  TeV, the
contribution can reach $\sim 20\%$ which is non-negligible.

\section{Conclusions}\label{Sec:Conclusion}

We have discussed the DM cascade decay induced by tiny soft
$C$-violating interactions in an extension of a left-right symmetric
model in which the DM particle is identified as a gauge singlet
scalar.  In this scenario, the DM particle may decay favorably into
leptonic final states through the intermediate $SU(2)_L$ triplets. We
have explored the parameter space which can well explain the current
PAMELA and Fermi-LAT or ATIC experimental data. It is shown that this
scenario predicts significant associating signals in the flux of high
energy neutrinos and gamma-rays, which is due to the correlated
couplings to neutrinos and the extra contributions from the charged
components of the triplet. We have found that the predicted
neutrino-induced muon flux is dominated by the contribution from the
neutrinos directly from the triplet decay. The gamma-ray radiation
from the charged triplet scalars through VIB process can be
significant in the case that the triplets are much lighter that the DM
particle.

\acknowledgments

This work is  supported in part by the National Basic Research Program of China (973
Program) under Grants No. 2010CB833000;  the National Nature 
Science Foundation of China (NSFC) under Grants No. 10975170, No. 10821504 
and No. 10905084; and the Project of Knowledge Innovation Program (PKIP) of 
the Chinese Academy of Science.

\end{document}